\documentclass[AMA,STIX1COL]{WileyNJD-v2}
\usepackage{moreverb}
\usepackage{bm}
\usepackage{hyperref}
\usepackage{verbatim}
\usepackage{url}
\theoremstyle{definition}
\newtheorem{exmp}{Example}[section]
\usepackage{tcolorbox}
\usepackage{graphicx}

\newcommand\BibTeX{{\rmfamily B\kern-.05em \textsc{i\kern-.025em b}\kern-.08em
T\kern-.1667em\lower.7ex\hbox{E}\kern-.125emX}}

\articletype{Article Type}%

\received{<day> <Month>, <year>}
\revised{<day> <Month>, <year>}
\accepted{<day> <Month>, <year>}

\makeatletter

\newcommand*{\addFileDependency}[1]{
\typeout{(#1)}
\@addtofilelist{#1}
\IfFileExists{#1}{}{\typeout{No file #1.}}
}\makeatother

\begin{document}

\title{Model-robust standardization in cluster-randomized trials}

\author[1,2]{Fan Li*}

\author[1,2]{Jiaqi Tong}

\author[1,2]{Xi Fang}

\author[3]{Chao Cheng}

\author[4]{Brennan C. Kahan}

\author[5]{Bingkai Wang}

\authormark{Li \textsc{et al.}}

\address[1]{Department of Biostatistics, Yale School of Public Health, New Haven, CT, USA}

\address[2]{Center for Methods in Implementation and Prevention Science, Yale School of Public Health, New Haven, CT, USA}

\address[3]{Department of Statistics and Data Science, Washington University, St. Louis, MO, USA}

\address[4]{MRC Clinical Trials Unit at UCL, London, UK}

\address[5]{Department of Biostatistics, School of Public Health, University of
Michigan, Ann Arbor, MI, USA}

\corres{Fan Li, Department of Biostatistics, Yale University School of Public Health, New Haven, Connecticut, USA.\\ \email{fan.f.li@yale.edu}}

\abstract[Abstract]{
In cluster-randomized trials, generalized linear mixed models and generalized estimating equations have conventionally been the default analytic methods for estimating the average treatment effect as routine practice. However, recent studies have demonstrated that their treatment effect coefficient estimators may correspond to ambiguous estimands when the models are misspecified or when there exists informative cluster sizes. In this article, we present a unified approach that standardizes output from a given regression model to ensure estimand-aligned inference for the treatment effect parameters in cluster-randomized trials. We introduce estimators for both the cluster-average and the individual-average treatment effects (marginal estimands) that are always consistent regardless of whether the specified working regression models align with the unknown data generating process. We further explore the use of a deletion-based jackknife variance estimator for inference. The development of our approach also motivates a natural test for informative cluster size. Extensive simulation experiments are designed to demonstrate the advantage of the proposed estimators under a variety of scenarios. The proposed model-robust standardization methods are implemented in the \texttt{MRStdCRT} R package.}


\jnlcitation{\cname{%
\author{Li F.}, 
\author{J. Tong}, 
\author{X. Fang}, 
\author{C. Cheng}, 
\author{B. Kahan}, and 
\author{B. Wang}} (\cyear{2025+}), 
\ctitle{Model-robust standardization in cluster-randomized trials}, \cjournal{Submitted to Statistics in Medicine}, \cvol{00:1--24}.}

\maketitle


\section{Introduction}\label{sec:introduction}
Cluster-randomized trials (CRTs) are experimental research designs that randomize interventions at the cluster or group level.\cite{murray1998design,turner2017review} All individual units within each cluster (such as a hospital, clinic, or worksite) receive the same treatment condition. CRTs have been widely employed in public health, biomedical sciences and social sciences, and are increasingly used for pragmatic clinical trials embedded in the healthcare systems.\cite{weinfurt2017pragmatic} A notable feature of a CRT is that the individual observations are nested within physical clusters. Due to this feature, the treatment effect can target changes in outcomes either at the cluster level or at the individual level. Kahan et al.\cite{kahan2023estimands,kahan2023demystifying} clarified that there could be different treatment effect estimands in CRTs. As two typical examples, the individual-average treatment effect (individual-ATE) answers the question ``\emph{what is the expected change in outcome associated with treatment among the population of individuals, regardless of cluster membership?}'', whereas the cluster-average treatment effect (cluster-ATE) answers the question ``\emph{what is the expected change in outcome associated with treatment among the population of clusters?}" In addition to their conceptual difference, these two estimands carry different magnitudes when there is informative cluster size (heterogeneity of outcomes or treatment effects by cluster size \cite{kahan2023informative}). The informative cluster size could arise due to inherent inter-cluster differences in experience or quality of care, or through systematic differences between participant baseline risk factors. In fact, the concept of treatment effect at different levels has appeared as early as in Donner and Klar,\cite{donner2000design} who have indicated in their textbook that:
\vspace{0.05cm}
\begin{tcolorbox}
``\emph{The target of inference in such studies (CRTs) could be at either the individual level or community level\ldots These examples show the importance of investigators explicitly formulating and stating the hypothesis under test in cluster randomization trials.}'' --- p.13 in Donner and Klar\cite{donner2000design}
\end{tcolorbox}
\noindent The CONSORT 2010 extension to CRTs has also mentioned an item to emphasize the level of the study objectives as:
\vspace{0.05cm}
\begin{tcolorbox}
``\emph{Descriptions of specific objectives and hypotheses need to make it clear whether they pertain to the individual participant level, the cluster level, or both. }'' --- Item 2b in Campbell et al.\cite{campbell2012consort}
\end{tcolorbox}
\noindent Relatedly, Imai et al.\citep{imai2009essential} outlined different estimands in pair-matched CRTs, depending on whether clusters and individuals within clusters are considered as fixed or randomly sampled from a larger population, but did not specifically address informative cluster size. Benitez et al.\cite{benitez2023defining} described estimands based on doubly-weighted potential outcomes at both the cluster and the individual levels. Under informative cluster size, Kahan et al.\cite{kahan2023demystifying} clarified estimands in CRTs by connecting the individual-ATE and cluster-ATE with marginal and cluster-specific quantities. 

Several prior studies have investigated estimators that target the cluster-ATE or individual-ATE as two marginal estimands in CRTs. Without baseline covariate adjustment, Table 3 in Kahan et al.\cite{kahan2023estimands} provided a recipe for consistent estimators using either the individual-level data or cluster-summary data; they recommended the (weighted) independence generalized estimating equations (GEE) when the cluster size is informative. Wang et al.\cite{wang2022two} pointed out that the probability limit of the treatment coefficient estimator from GEE with an exchangeable working correlation structure is neither the cluster-ATE nor the individual-ATE. Similar findings also apply to parallel-arm CRTs with a baseline period.\cite{lee2025should} With covariate adjustment, Su and Ding \cite{su2021model} established the asymptotic theory for a class of linearly-adjusted analysis of covariance estimators for estimating the weighted average treatment effects in CRTs, assuming working independence. Balzer et al.\cite{balzer2019new,balzer2023two} and Benitez et al.\cite{benitez2023defining} studied targeted maximum likelihood estimators with machine learning methods for covariate adjustment in CRTs, and extended their methods to address missing outcomes;\cite{nugent2024blurring} A further extension allowing for covariates missing not at random is given in Wang et al.\cite{wang2024handling} Wang et al.\cite{wang2024model} developed efficient covariate-adjusted estimators for estimating both the cluster-ATE and individual-ATE, allowing for informative within-cluster subsampling. Despite these recent developments, the most common practice for analyzing CRTs relies on fitting regression models suitable for clustered data, including the (generalized) linear mixed models and GEE. For example, Fiero et al.\cite{fiero2016statistical} reviewed 86 CRTs published between August 2013 and July 2014 and found that 52\% used mixed models and 16\% used GEE in their primary analyses. In a review of 100 CRTs from the United Kingdom’s National Institute for Health Research online Journals Library (from January 1997 to July 2021), 80\% of the studies adopted a linear mixed or generalized linear mixed model for data analysis.\citep{offorha2022statistical} For practitioners, guidance has been limited on how to draw robust inference about the cluster-ATE and the individual-ATE using common regression models for clustered data, irrespective of choices regarding covariate adjustment, working correlation and random-effects specification.

Focusing on marginal estimands, our primary objective is to describe a unified solution to achieve estimand-aligned analysis of CRTs based on common regression models for clustered data. We explain how one can standardize the output of any fitted regression model (i.e. by taking the predicted outcome values from a fitted model and averaging in a manner aligned with the target estimand) to target the cluster-ATE and the individual-ATE in CRTs. Because the treatment effect coefficients from linear mixed models and exchangeable GEE may be biased in the presence of informative cluster size, independence GEE (with or without inverse cluster size weights depending on the target estimand) has been recommended as a reliable alternative. However, despite its simplicity, independence GEE may not be a natural or preferred choice for the analysis of CRTs as it does not explicitly model the correlation structure. In contrast, our model-robust standardization approach allows one to use a much wider class of regression models for clustered data beyond independence GEE (e.g., allowing the use of linear mixed models and exchangeable GEE), and still achieve consistent estimation of the cluster-ATE and individual-ATE. Although the standardization procedure appears model-based, its construction ensures no asymptotic bias to the target estimand even though the fitted regression model does not perfectly align with the unknown data generating process; hence the standardization procedure is in fact model-robust. In summary, the model-robust standardization procedure bridges the classical literature for model-based analysis of CRTs and a recent strand of developments for estimand-aligned inference in CRTs.

The remainder of this article is organized as follows. Section \ref{sec:estimands} formalizes the treatment effect estimands in CRTs. Section 3 provides the construction of the standardization estimator for both the cluster-ATE and the individual-ATE, discusses implementation strategies and statistical inference by the jackknife. A test for informative cluster size is also provided as a direct by-product of the standardization estimator. Section 4 reports a series of simulations to illustrate the properties of the standardization estimator. Section 5 provides a reanalysis of a completed CRT to illustrate the proposed standardization estimators, and Section 6 provides a discussion about the implications of this work and several future research directions.

\section{Formalizing the treatment effect estimands}\label{sec:estimands}
We consider a standard parallel-arm CRT, where individuals are nested within clusters. We assume that data observations are collected through $m$ clusters, each of which contains $N_{i}$ individuals that we include in a study. We assume the cluster size $N_{i}$ is the study population of interest in each cluster and no further within-cluster subsampling is considered; methods that address the generalization to potentially unobserved cluster population can be found in Wang et al.\cite{wang2024model} and may require the additional knowledge of the source population size in each cluster. We let $A_i\in\{0,1\}$ be the randomized cluster-level treatment indicator, with $A_i=1$ indicating the assignment to the treatment condition and $A_i=0$ to usual care. Under the potential outcomes framework, we define $\{Y_{ij}(1),Y_{ij}(0)\}$ as a pair of potential outcomes for each individual $j \in \{1,\dots, N_i\}$ under the treatment and usual care conditions, respectively. Writing $f(a,b)$ as a pre-specified contrast function, we define a general class of weighted average treatment effect in CRTs as 
$$\Delta_{\omega}=f(\mu_\omega(1),\mu_\omega(0)),$$
where the weighted average potential outcome under treatment condition $A_i=a$ is
\begin{align}\label{eq:wate}
\mu_\omega(a)=\frac{E\left((\omega_i/N_i)\sum_{j=1}^{N_i}Y_{ij}(a)\right)}{E(\omega_i)}.
\end{align}
Here, $\omega_i$ is a pre-specified cluster-specific weight determining the contribution of each cluster to the target estimand, and is allowed to be a function of the cluster size $N_i$, or additional cluster-level covariates $\bm{H}_i$. In this formulation, the expectation is taken with respect to the super-population of clusters, from which the observed clusters are considered as a random sample. In CRTs, two typical marginal estimands of interest arise. First, setting $\omega_i=1$ gives each cluster equal weight and leads to the \emph{cluster-average treatment effect}, $\Delta_C=f(\mu_C(1),\mu_C(0))$, with
\begin{align}\label{eq:c-ate}
\mu_C(a)=E\left(\frac{\sum_{j=1}^{N_i}Y_{ij}(a)}{N_i}\right).
\end{align}
The cluster-ATE addresses the expected change in outcome associated with treatment for the population of clusters and is a natural target for interventions that are designed to improve, for example, provider practice for patient care. 
Second, setting $\omega_i=N_i$ gives equal weight to each individual in the study regardless of their cluster membership and leads to the \emph{individual-average treatment effect} (also referred to participant-average treatment effect\cite{kahan2023demystifying}), $\Delta_I=f(\mu_I(1),\mu_I(0))$, with
\begin{align}\label{eq:i-ate}
\mu_I(a)=\frac{E\left(\sum_{j=1}^{N_i}Y_{ij}(a)\right)}{E(N_i)},
\end{align}
Although estimand \eqref{eq:i-ate} does not appear to be an individual-average quantity from a first look, its finite-population analogue can be defined as $\sum_{i=1}^m\sum_{j=1}^{N_i}Y_{ij}(a)/\sum_{i=1}^m N_i$ (by replacing the population-level expectation with an empirical one)\cite{su2021model} and clearly reveals its feature. The individual-ATE addresses the expected change in outcome associated with treatment across the population of individuals across clusters, and can be relevant when the intervention is targeting either the cluster or the individual (despite cluster randomization).

To help distinguish between these two estimands, we offer three desiderata:
\begin{enumerate}
\item (\emph{Conceptual difference}) The cluster-ATE and the individual-ATE estimands focus on different units of inference,\cite{hemming2023commentary} regardless of unit of data analysis. The former addresses changes at the cluster level (either based on a cluster-level outcome or aggregated individual-level outcomes) and equally weights each cluster regardless of cluster size; the latter addresses changes for the population of individuals pooled across clusters and will upweight larger clusters. 
\item (\emph{Rule of thumbs}) The choice between these two estimands should be driven by the scientific question,\cite{kahan2023estimands} and depends on whether the interest lies in understanding the effect of the intervention on the clusters themselves, or on the individuals. Typically, the cluster-ATE may be of direct interest if a cluster-level intervention is studied to improve provider performance; thus the cluster-ATE is unique to CRTs and less relevant to individually randomized trials without an explicit cluster structure. On the other hand, the individual-ATE is likely of direct interest, if this is an individual-level intervention is studied but cluster randomization is chosen due to logistical and administrative reasons (or if the intent is to mimic an individually-randomized trial). Thus, the individual-ATE resembles the conventional estimand that one would have estimated in an individually randomized trial. Despite the rule of thumbs, we do not rule out the possibility where both estimands can be of interest.
\item (\emph{Mathematical difference}) The cluster-ATE and individual-ATE estimands will be different in magnitude when the cluster size is informative, i.e., when the cluster size is marginally associated with the outcome or the treatment contrast.\cite{kahan2023estimands,kahan2023informative} If the cluster size is non-informative and individual outcomes are marginally identically distributed within a cluster,$\mu_C(a)=\mu_I(a)$ and $\Delta_C=\Delta_I$ and these two estimands coincide.\cite{wang2021mixed}
\end{enumerate}

Beyond these two estimands, one may also choose $\omega_i$ as an explicit function to emphasize a target population with certain cluster attributes. For example, it is possible to specify larger $\omega_i$ for urban hospitals and smaller $\omega_i$ for rural hospitals if there is a scientific rationale to target a population with a higher proportion of urban hospitals. In addition, for a single binary cluster covariate $H_i\in\{0,1\}$, one can choose $\omega_i=I(H_i=1)$ to arrive at a subgroup cluster-average treatment effect estimand, $\mu_{\text{C}}(a|1)=E\left({\sum_{j=1}^{N_i}Y_{ij}(a)}/{N_i}\Big|H_i=1\right)$. A similar subgroup estimand can be obtained from $\omega_i=I(H_i=0)$. While we will focus on the cluster-ATE and individual-ATE, the model-robust standardization approach below applies to general choices of $\omega_i$ as any pre-specified functions of $N_i$ and $\bm{H}_i$.

\section{Model-robust Standardization based on Regression Models for CRTs}
\subsection{Constructing the estimator}
In a CRT, it is impossible to fully observe $\{Y_{ij}(1),Y_{ij}(0)\}$ for each individual, so that the estimand $\mu_{\omega}(a)$ is not directly estimable. We describe two assumptions that are plausible in CRTs. First, the observed outcome $Y_{ij}$, under the cluster-level Stable Unit Treatment Value Assumption, can be expressed as $Y_{ij}=A_iY_{ij}(1)+(1-A_i)Y_{ij}(0)$. Second, we assume cluster-level randomization such that $A_i\perp\{\bm{Y}_i(1),\bm{Y}_i(0)\}|\{\bm{X}_i,\bm{H}_i,N_i\}$, where, for cluster $i$, $\bm{Y}_i(a)=\left(Y_{i1}(a),\ldots,Y_{iN_i}(a)\right)^T$, $\bm{X}_i=\left(\bm{X}_{i1},\ldots,\bm{X}_{iN_i}\right)^T$, $\bm{H}_i$ is the collection of cluster-level covariates defined earlier, and $N_i$ is the cluster size. The randomization assumption holds by design, and includes as special cases (R1) simple randomization (where cluster-level randomization is determined by flipping a coin), (R2) stratified randomization\cite{lewsey2004comparing} (where one or more discrete cluster-level factor is considered to stratify clusters prior to randomization); (R3) pair-matching\cite{imai2009essential} (where randomization is done within pairs of clustered formed based on similar cluster-level attributes) and (R4) covariate-constrained randomization\cite{li2016evaluation,li2017evaluation} (where randomization is conducted within the subspace where sufficient balance is achieved according to baseline covariates). Under randomization schemes (R2)-(R4), the randomization of $A_i$ is conducted within levels of baseline covariates and the randomization probability may or may not vary according to levels of baseline covariates. 

Under this general setup, the model-robust standardization estimator for $\Delta_W$ takes the form of 
$$\widehat{\Delta}_\omega=f(\widehat{\mu}_\omega(1),\widehat{\mu}_\omega(0)),$$
with the average potential outcomes estimated by 
\begin{align}\label{eq:EST}
\widehat{\mu}_\omega(a)=\sum_{i=1}^m \frac{\omega_i}{\omega_{+}}\left\{\underbrace{\widehat{E}(\overline{Y}_{i}|A_i=a,\bm{X}_i,\bm{H}_i,N_i)}_{\text{regression prediction}}+\underbrace{\frac{I(A_i=a)\left(\overline{Y}_i-\widehat{E}(\overline{Y}_{i}|A_i=a,\bm{X}_i,\bm{H}_i,N_i)\right)}{\pi(\bm{X}_i,\bm{H}_i,N_i)^a\left(1-\pi(\bm{X}_i,\bm{H}_i,N_i)\right)^{1-a}}}_{\text{weighted cluster-level residual}}\right\},
\end{align}
where $\omega_{+}=\sum_{i=1}^m \omega_i$ is the sum of weights across clusters, $\pi(\bm{X}_i,\bm{H}_i,N_i)=P(A_i=a|\bm{X}_i,\bm{H}_i,N_i)$ is the randomization probability of each cluster (see Section \ref{ss:RandomProb}), and $\widehat{E}(\overline{Y}_{i}|A_i=a,\bm{X}_i,\bm{H}_i,N_i)$ is the conditional mean of the cluster average outcome, $\overline{Y_i}=N_i^{-1}\sum_{i=1}^{N_i}Y_{ij}$, possibly adjusting for baseline covariates and cluster size; this could be estimated via any outcome regression model (see Section \ref{sec:OR_estimator}). We first offer two remarks about \eqref{eq:EST} before discussing model specifications.

First, $\widehat{\mu}_\omega(a)$ is a weighted average of the cluster-specific contributions, with the weight proportional to $\omega_i$. Each cluster-specific contribution takes the form of a regression prediction $\widehat{E}(\overline{Y}_{i}|A_i=a,\bm{X}_i,\bm{H}_i,N_i)$, plus a weighted cluster-level residual. This is different from a purely model-based standardization estimator, $\sum_{i=1}^m \frac{\omega_i}{\omega_{+}}\widehat{E}(\overline{Y}_{i}|A_i=a,\bm{X}_i,\bm{H}_i,N_i)$, which is consistent only when the regression model is correctly specified.\cite{zou2009assessment} Due to the presence of a weighted residual term in \eqref{eq:EST}, $\widehat{\mu}_W(a)$ is consistent and asymptotically normal for estimating $\Delta_\omega$, regardless of whether the outcome regression model for the cluster mean outcome $\widehat{E}(\overline{Y}_{i}|A_i=a,\bm{X}_i,\bm{H}_i,N_i)$ matches the true data generating process (Web Appendix A). This property then offers sufficient flexibility in model specification for $\widehat{E}(\overline{Y}_{i}|A_i=a,\bm{X}_i,\bm{H}_i,N_i)$. In other words, one can use \eqref{eq:EST} as a unified solution to standardize the output from any fitted regression model for estimand-aligned inference in CRTs. By standardizing the regression model output, we mean taking the predicted outcome values from a fitted model and averaging in a manner aligned with the target estimand; Section \ref{sec:OR_estimator} offers concrete examples.

Second, we show in the Web Appendix A that $\widehat{\mu}_\omega(a)$ is the solution to the unbiased estimating equations motivated by the efficient influence function\cite{van2000asymptotic,hines2022demystifying} of $\mu_{\omega}(a)$. This implies that, if the outcome model for $\widehat{E}(\overline{Y}_{i}|A_i=a,\bm{X}_i,\bm{H}_i,N_i)$ matches the true data generating process, then $\widehat{\mu}_\omega(a)$ will be fully efficient as the large-sample variance of $\widehat{\mu}_\omega(a)$ achieves the semiparametric efficiency lower bound of $\mu_\omega(a)$. Such an estimator can substantially improve the efficiency over the unadjusted two-sample weighted difference-in-means estimator:
\begin{align}\label{eq:two-sample}
\widehat{\mu}_{\omega}^{\text{unadj}}(a)=\sum_{i=1}^m \frac{\omega_i}{\omega_{+}} \left\{\frac{I(A_i=a)}{\pi(\bm{X}_i,\bm{H}_i,N_i)^a\left(1-\pi(\bm{X}_i,\bm{H}_i,N_i)\right)^{1-a}}\right\}\overline{Y}_i,
\end{align}
This unadjusted estimator can also be obtained by setting $\widehat{E}(\overline{Y}_{i}|A_i=a,\bm{X}_i,\bm{H}_i,N_i)=0$ in \eqref{eq:EST}. For completeness, we provide the explicit forms of the model-robust standardization estimator for the cluster-ATE and individual-ATE as
\begin{align}\label{eq:c-ATE-estimator}
\widehat{\mu}_C(a)=\frac{1}{m}\sum_{i=1}^m \left\{\widehat{E}(\overline{Y}_{i}|A_i=a,\bm{X}_i,\bm{H}_i,N_i)+\frac{I(A_i=a)\left(\overline{Y}_i-\widehat{E}(\overline{Y}_{i}|A_i=a,\bm{X}_i,\bm{H}_i,N_i)\right)}{\pi(\bm{X}_i,\bm{H}_i,N_i)^a\left(1-\pi(\bm{X}_i,\bm{H}_i,N_i)\right)^{1-a}}\right\},
\end{align}
\begin{align}\label{eq:i-ATE-estimator}
\widehat{\mu}_I(a)=\sum_{i=1}^m \frac{N_i}{N}\left\{\widehat{E}(\overline{Y}_{i}|A_i=a,\bm{X}_i,\bm{H}_i,N_i)+\frac{I(A_i=a)\left(\overline{Y}_i-\widehat{E}(\overline{Y}_{i}|A_i=a,\bm{X}_i,\bm{H}_i,N_i)\right)}{\pi(\bm{X}_i,\bm{H}_i,N_i)^a\left(1-\pi(\bm{X}_i,\bm{H}_i,N_i)\right)^{1-a}}\right\},
\end{align}
where $N=\sum_{i=1}^m N_{i}$ is the total sample size.

\subsection{Specifying the randomization probability}\label{ss:RandomProb}
To implement the model-robust standardization estimator, one first needs to specify the cluster randomization probability, $\pi(\bm{X}_i,\bm{H}_i,N_i)$. This could be done under different randomization designs commonly seen in CRTs. Under simple randomization, the randomization probability is a constant, such that $\pi(\bm{X}_i,\bm{H}_i,N_i)=\pi\in(0,1)$. Under 1:1 randomization, one can simply use $\pi=1/2$. This same specification applies to stratified randomization and pair-matched randomization, with the requirement that stratification factors and matching variables (typically cluster-level variables) are captured in the set $\{\bm{X}_i,\bm{H}_i,N_i\}$. Under stratified randomization, it is most common to have the same randomization probability across strata, in which case $\pi(\bm{X}_i,\bm{H}_i,N_i)=\pi$, but stratum-specific randomization probability can be used for $\pi(\bm{X}_i,\bm{H}_i,N_i)$ whenever applicable. Under pair-matching, $\pi(\bm{X}_i,\bm{H}_i,N_i)=1/2$ as one cluster is randomly selected to receive intervention from a pair of two clusters. 

Li et al.\citep{li2016evaluation,li2017evaluation} demonstrated that covariate-constrained randomization is an effective randomization method to improve baseline balance in CRTs. As a recap, covariate-constrained randomization involves the following steps. First, one specifies key cluster-level attributes that need to be balanced by randomization. Second, one characterizes a balance metric that defines whether a proposed randomization scheme is acceptable in terms of baseline balance between groups. Third, one either enumerates (when the number of clusters is relatively limited) all or simulates (when there is a large number of clusters) a large number of potential randomization schemes based on the available clusters; in the case of simulated randomization schemes, the duplicate schemes are removed. Finally, one determines the threshold for acceptable randomization schemes, and choose a constrained randomization space as a subset of schemes where sufficient covariate balance is achieved according to the threshold. To specify $\pi(\bm{X}_i,\bm{H}_i,N_i)$ under covariate-constrained randomization, we first let $R$ represent the number of all possible randomization schemes or the number of simulated unique schemes under constrained randomization. We let $m_1$, $m_0$ be the number of clusters assigned to treatment and control arms, respectively, and $m=m_1+m_0$. We then specify the $R\times m$ matrix $\bm{T}$, so that each row $[\bm{T}]_{r\cdot}=(t_{r1},t_{r2},\ldots,t_{rn})$ indicates a unique randomization scheme, where the $(r,i)$th element $t_{ri}$ is $1$ if cluster $i$ is randomized to intervention according to the $r$th scheme and $0$ otherwise. Based on this constrained randomization matrix, one can approximate the vector of the randomization probabilities with $\left\{\pi(\bm{X}_1,\bm{H}_1,N_1),\pi(\bm{X}_2,\bm{H}_2,N_2),\ldots,\pi(\bm{X}_m,\bm{H}_m,N_m)\right\}=\bm{1}_R^T\bm{T}/R$. In the special case that $m/2$ clusters are randomized to each condition and the balance metric in a symmetric quadratic form (such as the Mahalanobis distance), it is often sufficient to use $\pi(\bm{X}_i,\bm{H}_i,N_i)=0.5$.

\subsection{Specifying the outcome regression model}\label{sec:OR_estimator}
In principle, we can compute $\widehat{E}(\overline{Y}_{i}|A_i=a,\bm{X}_i,\bm{H}_i,N_i)$ using any parametric or semiparametric regression models designed for clustered data. We give several familiar examples on how this term might be obtained. We present models by adjusting for all covariates included in $\{\bm{X}_i,\bm{H}_i,N_i\}$; this is not necessary, as the consistency of our standardization estimator does not have stringent requirement on any of these outcome regression models. For example, one could omit the adjustment for cluster size or any other baseline covariates without affecting consistency. In practice, we carry the same recommendations made for individually randomized trials that all baseline covariates that adjusted for in the analysis should be pre-specified and include those that are prognostic of the outcome.\citep{zeng2021propensity,wang2023model} Finally, in all of the examples presented below, we consider only the main effects as linear terms (akin to the ANCOVA1 specification in Wang et al.\cite{wang2021mixed}), but more flexible regression models including higher-order terms and interaction terms can be easily specified with minor modification. 

\begin{exmp}[\emph{Linear model or generalized linear model} based on cluster-level summaries]\label{example1}
Because ${E}(\overline{Y}_{i}|A_i=a,\bm{X}_i,\bm{H}_i,N_i)$ is a cluster-level expectation of the mean outcome, a natural approach is to consider a cluster-level mean regression:
\begin{align}\label{model:lm}
\overline{Y}_{i}=\beta_0+\beta_1 A_i + \bm{\beta}_2^T \overline{\bm X}_i+\bm{\beta}_3^T {\bm H}_i+\beta_4 N_i+\epsilon_i,
\end{align}
where $\overline{\bm X}_i=N_i^{-1}\sum_{j=1}^{N_i}\bm{X}_{ij}$ is the cluster-level mean covariate, and a working assumption is that the error $\epsilon_i$ follows a mean-zero normal distribution with a constant variance. This working assumption need not be correct (as $\overline{Y}_i$ is often heteroscedastic with variable cluster sizes) and is primarily to facilitate the ordinary least squares estimation. Notably, the cluster-level linear regression can be easily operationalized even if $Y_{ij}$ is binary or count, as the error assumption is only a working device. More generally, this approach has been known as the two-stage approach in the analysis of CRTs.\citep{johnson2015recommendations,hayes2017cluster} From \eqref{model:lm}, it is immediate that 
$$\widehat{E}(\overline{Y}_{i}|A_i=a,\bm{X}_i,\bm{H}_i,N_i)=\widehat\beta_0+\widehat\beta_1 A_i + \widehat{\bm{\beta}}_2^T \overline{\bm X}_i+\widehat{\bm{\beta}}_3^T {\bm H}_i+\widehat\beta_4 N_i,$$ 
which can be used in the model-robust standardization estimators \eqref{eq:c-ATE-estimator} and \eqref{eq:i-ATE-estimator}. Alternatively, when the outcome is binary, one could also use the generalized linear model (GLM) for cluster-level proportions. For example, we extend the unadjusted logistic regression in Kahan et al.\cite{kahan2023demystifying} with baseline adjustment as:
\begin{align}\label{model:glm}
\text{logit}({E}(\overline{Y}_{i}|A_i=a,\bm{X}_i,\bm{H}_i,N_i))=\lambda_0+\lambda_1 A_i + \bm{\lambda}_2^T \overline{\bm X}_i+\bm{\lambda}_3^T {\bm H}_i+\lambda_4 N_i,
\end{align}
where the Gaussian working variance function can be used for the cluster-level proportions. Model \eqref{model:glm} then implies that $\widehat{E}(\overline{Y}_{i}|A_i=a,\bm{X}_i,\bm{H}_i,N_i)=\text{expit}(\widehat\lambda_0+\widehat\lambda_1 A_i + \widehat{\bm{\lambda}}_2^T \overline{\bm X}_i+\widehat{\bm{\lambda}}_3^T {\bm H}_i+\widehat\lambda_4 N_i)$ with $\text{expit}(x)=(1+e^{-x})^{-1}$. 

The merit of this cluster-level regression is that it is easy to implement (for example, using the \texttt{lm} or \texttt{glm} function in \texttt{R}). The potential drawback is that, with a limited number of clusters that are frequently seen in CRTs, there is a cap in the cluster-level degrees of freedom that limits the number of cluster-level covariates to be adjusted for.  
\end{exmp}

\begin{exmp}[\emph{Linear mixed model based on individual-level observations}]\label{example2} 
As a dominating approach in analyzing CRTs, the linear mixed model fitted with individual-level observations\citep{wang2021mixed} can be represented by
\begin{align}\label{model:lmm}
Y_{ij}=\gamma_0+\gamma_1 A_i + \bm{\gamma}_2^T(\bm{X}_{ij}-\overline{\bm X}_i)+\bm{\gamma}_3^T \overline{\bm X}_i+\bm{\gamma}_4^T {\bm H}_i+\gamma_5 N_i+b_i+\epsilon_{ij},
\end{align}
where $\bm{\gamma}_2$ and $\bm{\gamma}_3$ separately account for the within-cluster and between-cluster effects of the individual-level covariates. In the analysis of clustered data, this model has been referred to as the contextual effect model; see, for example, Raudenbush\cite{raudenbush1997statistical} and Tong et al.\cite{tong2022accounting} Setting $\bm{\gamma}_2=\bm{\gamma}_3$ eliminates the contextual effect in the model formulation. In \eqref{model:lmm}, $b_i\sim \mathcal{N}(0,\sigma_b^2)$ is the cluster-level random intercept and $\epsilon_{ij}\sim \mathcal{N}(0,\sigma_\epsilon^2)$ is the random error. The outcome ICC given treatment and covariates, implied by the linear mixed model, is $\rho=\sigma_b^2/(\sigma_b^2+\sigma_\epsilon^2)$. This quantity is of great interest in CRTs and plays a critical role in determining the sample size and power.\cite{turner2017review} Per the CONSORT extension of CRTs,\cite{campbell2012consort} reporting the ICC values in completed trials may facilitate the planning of future studies with a similar endpoint and has been a recommended practice. As in Example \ref{example1}, the linear mixed model can be fit to binary and count data because the distributional assumptions for the random effect and the error term are only working assumptions; we refer to Wang et al.\cite{wang2021mixed} for a rigorous treatment of properties of the linear mixed models for the analysis of CRTs without informative cluster size. Under model \eqref{model:lmm}, we have
\begin{align*}
\widehat{E}(\overline{Y}_{i}|A_i=a,\bm{X}_i,\bm{H}_i,N_i)=
\widehat\gamma_0+\widehat\gamma_1 A_i +\widehat{\bm{\gamma}}_3^T \overline{\bm X}_i+\widehat{\bm{\gamma}}_4^T {\bm H}_i+\widehat{\gamma_5} N_i,
\end{align*}
where the coefficient estimator can be obtained by maximum likelihood or restricted maximum likelihood, available from, for example, the \texttt{lmer} function in the \texttt{lme4} \texttt{R} package. Compared to the cluster-level models in Example \ref{example1}, the linear mixed model can adjust for individual-level covariates and provides an estimate of the ICC.
\end{exmp}

\begin{exmp}[\emph{Generalized linear mixed model based on individual-level observations}]\label{example3}
With a binary or count outcome, generalized linear mixed models have been more commonly used along with a canonical link function in CRTs. Assuming a binary outcome, a logistic mixed model can be written as 
\begin{align}\label{model:logit-lmm}
\text{logit}\{P(Y_{ij}=1|A_i,\bm{X}_i,\bm{H}_i,N_i,c_i)\}=\alpha_0+\alpha_1 A_i + \bm{\alpha}_2^T(\bm{X}_{ij}-\overline{\bm X}_i)+\bm{\alpha}_3^T \overline{\bm X}_i+\bm{\alpha}_4^T {\bm H}_i+\alpha_5 N_i+c_i,
\end{align}
where $c_i\sim \mathcal{N}(0,\sigma_c^2)$ is an independently distributed cluster-level random intercept. A frequent practice is to consider $\alpha_1$ as the treatment effect, but such a definition is grounded on model \eqref{model:logit-lmm} and is only a clear conditional average treatment effect parameter on the log odds ratio scale when model \eqref{model:logit-lmm} is correctly specified. If the fitted logistic mixed model is incorrect, $\alpha_1$ in general does not have a straightforward interpretation as a treatment effect estimand defined under the potential outcomes framework. However, this does not preclude us from using \eqref{model:logit-lmm} to derive a model-robust standardization estimator. Specifically, the induced cluster-level mean adjusting for covariates can be computed as
\begin{align}
\widehat{E}(\overline{Y}_{i}|A_i= a,\bm{X}_i,\bm{H}_i,N_i)&=\frac{1}{N_i}\sum_{i=1}^{N_i}\widehat{E}(Y_{ij}|A_i=a,\bm{X}_i,\bm{H}_i,N_i)\nonumber\\
&=\frac{1}{N_i}\sum_{i=1}^{N_i}\int_0^\infty \text{expit}\left\{\widehat\alpha_0+\widehat\alpha_1 A_i + \widehat{\bm{\alpha}}_2^T(\bm{X}_{ij}-\overline{\bm X}_i)+\widehat{\bm{\alpha}}_3^T \overline{\bm X}_i+\widehat{\bm{\alpha}}_4^T {\bm H}_i+\widehat\alpha_5 N_i+c\right\}\phi(c|0,\widehat{\sigma}_c^2)dc,\label{eq:logit-normal}
\end{align}
where \(\phi(.)\) denotes the distribution of the random effect \(c\). The last expression \eqref{eq:logit-normal} is essentially a logistic-normal integral that can either be numerically approximated with arbitrary accuracy (see, for example, Crouch and Spiegelman\cite{crouch1990evaluation}), or evaluated using Monte Carlo approximation. Since typically only a single random intercept is considered, a close approximation of \eqref{eq:logit-normal} is given by the simpler expression ($\Pi$ is the mathematical constant)\cite{hedeker2018note}
\begin{align*}
&\widehat{E}(\overline{Y}_{i}|A_i=a,\bm{X}_i,\bm{H}_i,N_i)\approx \frac{1}{N_i}\sum_{i=1}^{N_i} \text{expit}\left\{\left(\widehat\alpha_0+\widehat\alpha_1 a + \widehat{\bm{\alpha}}_2^T(\bm{X}_{ij}-\overline{\bm X}_i)+\widehat{\bm{\alpha}}_3^T \overline{\bm X}_i+\widehat{\bm{\alpha}}_4^T {\bm H}_i+\widehat\alpha_5 N_i\right)/\sqrt{\frac{\widehat{\sigma}_c^2+\Pi^2/3}{\Pi^2/3}}\right\},
\end{align*}
which obviates the need for numerical integration. Next, when a log linear mixed model is considered (for either a binary or count outcome), the working model can be expressed as
\begin{align}\label{model:log-lmm}
\text{log}\{E(Y_{ij}|A_i,\bm{X}_i,\bm{H}_i,N_i,d_i)\}=\xi_0+\xi_1 A_i + \bm{\xi}_2^T(\bm{X}_{ij}-\overline{\bm X}_i)+\bm{\xi}_3^T \overline{\bm X}_i+\bm{\xi}_4^T {\bm H}_i+\xi_5 N_i+d_i,
\end{align}
where $d_i\sim \mathcal{N}(0,\sigma_d^2)$. Due to collapsibility, we can compute
\begin{align*}
\widehat{E}(\overline{Y}_{i}|A_i=a,\bm{X}_i,\bm{H}_i,N_i)
=\frac{1}{N_i}\sum_{i=1}^{N_i} \exp\left\{\left(\widehat\xi_0+\frac{\widehat\sigma_d^2}{2}\right)+\widehat\xi_1 a + \widehat{\bm{\xi}}_2^T(\bm{X}_{ij}-\overline{\bm X}_i)+\widehat{\bm{\xi}}_3^T \overline{\bm X}_i+\widehat{\bm{\xi}}_4^T {\bm H}_i+\widehat\xi_5 N_i\right\}.
\end{align*}
That is, in log linear mixed models, marginalization over the random intercept only implies an overall intercept shift.\cite{ritz2004equivalence} The coefficient estimator for each generalized linear mixed model can be obtained from maximizing the approximate likelihood, through either the approach of penalized quasi-likelihood, Laplace approximation or adaptive Gaussian-quadrature;\cite{li2017evaluation} all three approaches have been implemented in the \texttt{glmer} function in the \texttt{lme4} \texttt{R} package.  
\end{exmp}

\begin{exmp}[\emph{Marginal models based on individual-level observations}]\label{example4}
Unlike the conditional models, marginal models obviate the need to condition on the latent random intercept. Instead, a generalized linear model is specified for the marginal mean (marginal in the sense that it is not a cluster-specific model with latent random effects), and a separate working correlation model is specified to account for the within-cluster associations between the outcomes.\cite{liang1986longitudinal} A general form of the marginal mean model is given by
\begin{align}\label{model:marginal}
g\{E(Y_{ij}|A_i,\bm{X}_i,\bm{H}_i,N_i)\}=\zeta_0+\zeta_1 A_i + \bm{\zeta}_2^T(\bm{X}_{ij}-\overline{\bm X}_i)+\bm{\zeta}_3^T \overline{\bm X}_i+\bm{\zeta}_4^T {\bm H}_i+\zeta_5 N_i,
\end{align}
and a general expression for the standardization step is therefore
\begin{align*}
\widehat{E}(\overline{Y}_{i}|A_i=a,\bm{X}_i,\bm{H}_i,N_i)
=\frac{1}{N_i}\sum_{i=1}^{N_i} g^{-1}\left\{\widehat\zeta_0+\widehat\zeta_1 a + \widehat{\bm{\zeta}}_2^T(\bm{X}_{ij}-\overline{\bm X}_i)+\widehat{\bm{\zeta}}_3^T \overline{\bm X}_i+\widehat{\bm{\zeta}}_4^T {\bm H}_i+\widehat\zeta_5 N_i\right\},
\end{align*}
which does not involve any integration regardless of the link function. The estimation routine for the regression coefficients is based on solving the GEE, and has been extensively studied for clustered data; see, Preisser et al.\cite{preisser2003integrated} for a detailed discussion on the merits of GEE for the design and analysis of parallel-arm CRTs. The implementation of GEE requires specifying the within-cluster correlation structure. Denoting $\bm{C}_i=\text{corr}(Y_{i1},\ldots,Y_{iN_i})$ as the working correlation structure for cluster $i$, the GEE regression estimator in CRTs can be based on either the independence working structure ($\bm{C}_i=\bm{I}_{N_i}$), exchangeable working structure ($\bm{C}_i=(1-\rho)\bm{I}_{N_i}+\rho\bm{1}_{N_i}\bm{1}^T_{N_i}$ with $\rho$ being the ICC), or arm-specific exchangeable working structure ($\bm{C}_i=\{A_i(1-\rho_1)+(1-A_i)(1-\rho_0)\})\bm{I}_{N_i}+\{A_i\rho_1+(1-A_i)\rho_0\}\bm{1}_{N_i}\bm{1}^T_{N_i}$ with $\rho_a$ being the arm-specific ICC).\cite{li2021sample} For a correctly specified marginal mean model, specifying an approximately correct working correlation structure generally leads to improved efficiency in estimating the regression coefficients compared to the independence working structure.\cite{li2021sample_bin} 
To improve the standard GEE implementation, Preisser et al.\cite{preisser2008finite} provided matrix-adjusted estimating equations (MAEE) to correct for the negative bias in the correlation estimates in the presence of a small number of clusters. Finally, more flexible specifications of the working correlation structure can be considered following the quadratic inference functions approach.\cite{westgate2012bias,yu2020evaluation} As a further example, the marginalized multilevel models exploit the strengths of the generalized linear mixed model (in Example \ref{example3}) to specify flexible within-cluster correlations but still enable the estimation of the marginal mean model parameters in \eqref{model:marginal}.\cite{heagerty2000marginalized} The tools for implementing marginal models are also developed. For instance, the general GEE routine can be implemented in the \texttt{geeglm} function in the \texttt{geepack} \texttt{R} package, whereas the MAEE routine can be found in the \texttt{geeCRT} \texttt{R} package. \texttt{R} code for implementing the quadratic inference function is available from the Appendix of Westgate.\cite{westgate2012bias} 
\end{exmp}

\subsection{Inference via cluster jackknifing}\label{sec:jack_var}
Because the model-robust standardization estimator can accommodate different specifications of the working outcome regression models, it is desirable to have a one-size-fits-all solution to statistical inference. Therefore, we consider a deletion-based cluster jackknife variance estimator that is simple to implement. The jackknife variance estimator has been previously studied for model-based inference with a small number of clusters. For example, in the context of linear regression of clustered data, many have shown that jackknife variance estimator often outperforms the bias-corrected cluster-robust variance estimators in terms of the test size in small samples.\cite{hansen2022jackknife} The jackknife variance estimator is also connected with the Mancl and DeRouen GEE variance estimator to combat finite-sample bias.\cite{mancl2001covariance,ouyang2024maintaining} Finally, compared to other resampling-based methods such as cluster bootstrap and permutation, the computation cost needed for jackknife is more affordable as one only needs to refit the model-robust standardization estimator at most $m$ times. 

To describe this estimator, for $g\in\{1,\ldots,m\}$, we define the leave-one-cluster-out estimator for $\mu_{\omega}(a)$ as
\begin{align}\label{LOCO}
\widehat{\mu}_\omega^{-g}(a)=\sum_{i\neq g}^m \frac{\omega_i}{\omega_{+}^{-g}}\left\{\widehat{E}^{-g}(\overline{Y}_{i}|A_i=a,\bm{X}_i,\bm{H}_i,N_i)+\frac{I(A_i=a)\left(\overline{Y}_i-\widehat{E}^{-g}(\overline{Y}_{i}|A_i=a,\bm{X}_i,\bm{H}_i,N_i)\right)}{\pi(\bm{X}_i,\bm{H}_i,N_i)^a\left(1-\pi(\bm{X}_i,\bm{H}_i,N_i)\right)^{1-a}}\right\},
\end{align}
where $\omega_{+}^{-g}\sum_{i\neq g} {\omega_i}$ is the sum of weights excluding cluster $g$. We further define the average of all deletion-based model-robust standardization estimators as $\overline{\mu}_\omega(a)=m^{-1}\sum_{g=1}^m \widehat{\mu}_\omega^{-g}(a)$. In \eqref{LOCO}, we compute the average potential outcome by leaving out cluster $g$ in the analytic sample; of note, the outcome regression model will also be fitted after leaving cluster $g$ out and we use $\widehat{E}^{-g}(\overline{Y}_{i}|A_i=a,\bm{X}_i,\bm{H}_i,N_i)$ to represent the refitted model. The jackknife variance estimator for the pair of the average potential outcome estimators is then given by
\begin{align}\label{JVE1}
\widehat{\bm{\Sigma}}=\frac{m-1}{m}\sum_{g=1}^m \begin{pmatrix}
\left\{\widehat{\mu}_\omega^{-g}(1)-\overline{\mu}_\omega(1)\right\}^2 & 
\left\{\widehat{\mu}_\omega^{-g}(1)-\overline{\mu}_\omega(1)\right\}\left\{\widehat{\mu}_\omega^{-g}(0)-\overline{\mu}_\omega(0)\right\}\\
\left\{\widehat{\mu}_\omega^{-g}(1)-\overline{\mu}_\omega(1)\right\}\left\{\widehat{\mu}_\omega^{-g}(0)-\overline{\mu}_\omega(0)\right\} & \left\{\widehat{\mu}_\omega^{-g}(0)-\overline{\mu}_\omega(0)\right\}^2
\end{pmatrix},
\end{align}
Hypothesis tests and confidence intervals can then be constructed with the $t(m-1)$ approximate sampling distribution.

\subsection{A model-robust test for informative cluster size}\label{sec:ICS_test}
The above development naturally suggests a statistical test for the presence of informative cluster size. Given that the cluster-ATE and individual-ATE are often of greater interest, we can test for the following null hypothesis
\begin{equation}\label{null_ICS}
    \mathcal{H}_0: f({\mu}_C(1),{\mu}_C(0))=f({\mu}_I(1),{\mu}_I(0)).
\end{equation}
Since informative cluster size results in $f({\mu}_C(1),{\mu}_C(0))\ne f({\mu}_I(1),{\mu}_I(0))$, rejecting $\mathcal{H}_0$ implies informative cluster size. A natural test statistic can be constructed as a contrast between $f(\widehat{\mu}_C(1),\widehat{\mu}_C(0))$ and $f(\widehat{\mu}_I(1),\widehat{\mu}_I(0))$ (up to some transformation), scaled by its variance similarly computed via cluster jackknifing. For estimands defined on the difference scale such that $f$ is the identity function, the null \eqref{null_ICS} is equivalent to testing whether the following:
\begin{equation}\label{null_ICS_original}
\mathcal{H}_0: {E\left(\left\{\frac{E(N_i)}{N_i}-1\right\}\left\{\sum_{j=1}^{N_i}\left(Y_{ij}(1)-Y_{ij}(0)\right)\right\}\right)}=0,
\end{equation}
which is given by the difference between the cluster-ATE and individual-ATE estimands. After some algebra, this null hypothesis can be re-expressed as
\begin{equation}\label{null_ICS_2}
\mathcal{H}_0: {\text{Cov}\left(N_i,\frac{1}{N_i}\left\{\sum_{j=1}^{N_i}\left(Y_{ij}(1)-Y_{ij}(0)\right)\right\}\right)}=0.
\end{equation}
Equation \eqref{null_ICS_2} reveals an important insight on a testable implication under informative cluster size. That is, informative cluster size will lead to a non-zero marginal covariance between the cluster size and the cluster-specific treatment effect. 

To proceed, when $f$ is the difference measure, we first compute the difference between the cluster-ATE and individual-ATE estimators as $\widehat{D} = \widehat{\Delta}_C - \widehat{\Delta}_I =(\widehat{\mu}_C(1)-\widehat{\mu}_C(0))-(\widehat{\mu}_I(1)-\widehat{\mu}_I(0))$. To estimate the variance of \( \widehat{D} \), we can use the jackknife resampling method described in Section \ref{sec:jack_var}. Specifically, we sequentially leave out one cluster at a time and recompute the difference estimators on the remaining \( m - 1 \) clusters to obtain leave-one-cluster-out estimators \( \widehat{D}^{(-i)} = \widehat{\Delta}_C^{(-i)} - \widehat{\Delta}_I^{(-i)} \) for each \( i = 1, \ldots, m \). The jackknife estimate of the variance of \( \widehat{D} \) is then given by 
$$\widehat{V}_{D} = \frac{m - 1}{m} \sum_{i=1}^m \left( \widehat{D}^{(-i)} - \overline{D} \right)^2,$$ 
where \( \overline{D} = \frac{1}{m} \sum_{i=1}^m \widehat{D}^{(-i)} \) is the average of the jackknife estimators. The final test statistic is constructed as \( T = \widehat{D} / \sqrt{\widehat{V}_D} \), which is approximately \( t \)-distributed with \( m - 1 \) degrees of freedom under the null hypothesis. This consists of the basis for computing p-values. The procedures for carrying out the test when $f$ defines the ratio effect follow a similar construction, except that one may define the test based on $\log(\widehat{\Delta}_C) - \log(\widehat{\Delta}_I)$ to improve numerical stability.

\section{Simulation experiments}\label{sec:simulation}
We conduct a series of simulation studies to compare model-robust standardization to standard practice of estimating regression coefficient. Throughout, we fix the number of simulations to $1000$ and focus on estimating $\Delta_C$ and $\Delta_I$; the choice of effect measure and $f$ depends on the type of the outcome and will be explained in each setting. In Sections \ref{s:sim;ss:NICS}-\ref{s:sim;ss:binary}, we investigate the 4 combinations arising from 2 outcome types (continuous or binary) and 2 scenarios for informative cluster sizes (absence or presence). Tables \ref{tab_NICS}-\ref{tab_binary_se_m100_cate} summarize the main results, and the corresponding Monte Carlo standard errors quantifying simulation uncertainty are given in Web Tables 1-4; the definitions of each uncertainty measure are given in Table 6 of Morris et al.\cite{morris2019using} Section \ref{sec:others} summarizes findings from additional simulations presented in the Web Appendix and Section \ref{sec:ics_test} illustrates the performance of the test for informative cluster size.

\subsection{Non-informative cluster size and a continuous outcome}\label{s:sim;ss:NICS}
We first compare the performance of the model-robust standardization estimator with the model-based treatment coefficient estimator under the setting of non-informative cluster size. We focus on a continuous outcome with two levels of sample sizes (number of clusters $m=30$ or $m=100$). We fix the expected total sample size such that $m  E(N_i)=3000$. The actual cluster sizes are drawn from a discrete uniform distribution supported between $20$ and $180$ ($m=30$) or between $6$ and $54$ ($m=100$). 

For the continuous outcome setting, we simulate one binary cluster-level covariate $H_{1i}\sim \text{Bernoulli}\{\Phi(\text{sin}(E(N_i)))\}$, where $\Phi$ is the standard normal cumulative distribution function, and one continuous cluster-level covariate $H_{2i}\sim \mathcal{N}(2+H_{1i}E(N_i)/10,9)$. We generate one continuous individual-level covariate $X_{1ij}\sim\mathcal{N}(H_{1i}H_{2i}+E(N_i)/100,16)$ and one binary individual-level covariate $X_{2ij}\sim \text{Bernoulli}\{\text{expit}(\text{log}(E(N_i))X_{1ij}H_{1i}+H_{2i})\}$.
We simulate the true potential outcomes from: 
\begin{align*}
    Y_{ij}(a)\sim \mathcal{N}\left(3+\frac{H_{1i}X_{1ij}^2}{5E\{N_i\}}+\text{cos}(H_{2i})X_{2ij}+|H_{2i}|\text{sin}(X_{2ij})-3a+\gamma_ia,1\right),~~~~a\in\{0,1\}.
\end{align*}
The random intercept $\gamma_i\sim\mathcal{N}(0,0.2)$ and induces a positive adjusted ICC $\approx 0.17$ under the treatment condition. In this case, we define the target estimands on the difference scale ($f(x,y)=x-y$), and analytically derive the cluster-ATE and individual-ATE as $-3$. We draw $A_i\sim\text{Bernoulli}(0.5)$, and set the observed outcome $Y_{ij}=A_iY_{ij}(1)+(1-A_i)Y_{ij}(0)$. Here, the true potential outcome model assumes a constant treatment effect across individuals but incorporates relatively nonlinear covariate functional forms. This design allows us to evaluate the performance of the model-robust standardization and model-based coefficient estimators under model misspecification, where the fitted models are allowed to differ from the true data generating mechanisms. Second, the cluster size \( N_i \) is neither associated with \( Y_{ij}(a) \) nor acts as an effect modifier. Consequently, there is no informative cluster size, ensuring equivalence between the cluster-ATE and the individual-ATE.

\begin{table}[ht!]
\caption{\label{tab_NICS} Simulation results under a data generating process with \textbf{non-informative} cluster size and a continuous outcome. LM: linear model based on cluster-level means; LMM: linear mixed model based on individual-level data; GEE-exch: marginal model fitted with GEE and an exchangeable working correlation structure; GEE-ind: marginal model fitted with GEE and working independence. Each empirical coverage probability is reported together with its Monte Carlo 95\% confidence limits (95\% CL). Additional Monte Carlo standard errors for performance metric estimates are presented in Web Table 1. }
\begin{center}
\resizebox{\textwidth}{!}{%
\begin{tabular}{cclcrrrrrr}
\toprule
\multirow{2}{*}{$m$} & Covariate & \multirow{2}{*}{Outcome model} & \multirow{2}{*}{Method}
& \multirow{2}{*}{BIAS} & MCSD & COV
& \multirow{2}{*}{BIAS} & MCSD & COV \\
& adjustment & & & & (AESE) & (95\% CL) & & (AESE) & (95\% CL)\\ \midrule
&&&& \multicolumn{3}{c}{$\Delta_C=-3$} & \multicolumn{3}{c}{$\Delta_I=-3$} \\
\multirow{16}{*}{$30$}
  & \multirow{8}{*}{$\times$}
      & \multirow{2}{*}{(W1) LM}
    & Coef & -0.6 & 1.72 (1.63) & 93.2 (91.8, 94.6) & \textbackslash & \textbackslash & \textbackslash \\
  & & & MRS & -0.6 & 1.72 (1.73) & 94.7 (93.5, 95.9) & -0.3 & 1.84 (1.90) & 93.8 (92.5, 95.1) \\
  & & \multirow{2}{*}{(W2) LMM}
    & Coef & -0.6 & 1.72 (1.63) & 93.2 (91.8, 94.6) & \textbackslash & \textbackslash & \textbackslash \\
  & & & MRS & -0.6 & 1.72 (1.73) & 94.7 (93.5, 95.9) & -0.3 & 1.84 (1.90) & 93.8 (92.5, 95.1) \\
  & & \multirow{2}{*}{(W3) GEE-exch}
    & Coef & -0.6 & 1.72 (1.76) & 95.1 (94.0, 96.2) & \textbackslash & \textbackslash & \textbackslash \\
  & & & MRS & -0.6 & 1.72 (1.73) & 94.7 (93.5, 95.9) & -0.3 & 1.84 (1.90) & 93.8 (92.5, 95.1) \\
  & & \multirow{2}{*}{(W4) GEE-ind}
    & Coef & -0.1 & 1.85 (1.95) & 94.4 (93.2, 95.6) & \textbackslash & \textbackslash & \textbackslash \\
  & & & MRS & -0.4 & 1.74 (1.74) & 94.7 (93.5, 95.9) & -0.1 & 1.85 (1.92) & 93.9 (92.6, 95.2) \\
[0.4ex]
  & \multirow{8}{*}{$\checkmark$}
      & \multirow{2}{*}{(W1) LM}
    & Coef & 0.2 & 0.30 (0.25) & 90.7 (89.1, 92.3) & \textbackslash & \textbackslash & \textbackslash \\
  & & & MRS & 0.2 & 0.30 (0.38) & 98.3 (97.5, 99.1) & 0.2 & 0.32 (0.40) & 97.6 (96.6, 98.6) \\
  & & \multirow{2}{*}{(W2) LMM}
    & Coef & 0.2 & 0.30 (0.25) & 90.8 (89.2, 92.4) & \textbackslash & \textbackslash & \textbackslash \\
  & & & MRS & 0.2 & 0.30 (0.35) & 97.2 (96.2, 98.2) & 0.2 & 0.32 (0.36) & 96.7 (95.6, 97.8) \\
  & & \multirow{2}{*}{(W3) GEE-exch}
    & Coef & 0.2 & 0.30 (0.35) & 97.6 (96.7, 98.5) & \textbackslash & \textbackslash & \textbackslash \\
  & & & MRS & 0.2 & 0.30 (0.35) & 97.2 (96.2, 98.2) & 0.2 & 0.32 (0.36) & 96.7 (95.6, 97.8) \\
  & & \multirow{2}{*}{(W4) GEE-ind}
    & Coef & 0.2 & 0.32 (0.38) & 96.7 (95.7, 97.7) & \textbackslash & \textbackslash & \textbackslash \\
  & & & MRS & 0.0 & 0.30 (0.36) & 97.0 (96.0, 98.0) & 0.2 & 0.32 (0.37) & 96.6 (95.5, 97.7) \\
[0.6ex]

&&&& \multicolumn{3}{c}{$\Delta_C=-3$} & \multicolumn{3}{c}{$\Delta_I=-3$} \\
\multirow{16}{*}{$100$}
  & \multirow{8}{*}{$\times$}
      & \multirow{2}{*}{(W1) LM}
    & Coef & -0.2 & 0.49 (0.47) & 94.8 (93.7, 95.9) & \textbackslash & \textbackslash & \textbackslash \\
  & & & MRS & -0.2 & 0.49 (0.48) & 95.3 (94.2, 96.4) & 0.0 & 0.54 (0.53) & 94.2 (93.0, 95.4) \\
  & & \multirow{2}{*}{(W2) LMM}
    & Coef & -0.2 & 0.49 (0.47) & 94.8 (93.7, 95.9) & \textbackslash & \textbackslash & \textbackslash \\
  & & & MRS & -0.2 & 0.49 (0.48) & 95.3 (94.2, 96.4) & 0.0 & 0.54 (0.53) & 94.2 (93.0, 95.4) \\
  & & \multirow{2}{*}{(W3) GEE-exch}
    & Coef & -0.2 & 0.49 (0.48) & 95.5 (94.5, 96.5) & \textbackslash & \textbackslash & \textbackslash \\
  & & & MRS & -0.2 & 0.49 (0.48) & 95.3 (94.2, 96.4) & 0.0 & 0.54 (0.53) & 94.2 (93.0, 95.4) \\
  & & \multirow{2}{*}{(W4) GEE-ind}
    & Coef & 0.0 & 0.54 (0.54) & 94.6 (93.5, 95.7) & \textbackslash & \textbackslash & \textbackslash \\
  & & & MRS & -0.2 & 0.49 (0.48) & 95.5 (94.5, 96.5) & 0.0 & 0.54 (0.53) & 94.3 (93.1, 95.5) \\
[0.4ex]
  & \multirow{8}{*}{$\checkmark$}
      & \multirow{2}{*}{(W1) LM}
    & Coef & 0.0 & 0.17 (0.16) & 94.2 (93.1, 95.3) & \textbackslash & \textbackslash & \textbackslash \\
  & & & MRS & 0.0 & 0.17 (0.18) & 96.9 (96.0, 97.8) & 0.0 & 0.18 (0.19) & 96.1 (95.1, 97.1) \\
  & & \multirow{2}{*}{(W2) LMM}
    & Coef & 0.0 & 0.17 (0.16) & 93.9 (92.8, 95.0) & \textbackslash & \textbackslash & \textbackslash \\
  & & & MRS & 0.0 & 0.17 (0.18) & 96.5 (95.6, 97.4) & 0.0 & 0.18 (0.19) & 95.8 (94.8, 96.8) \\
  & & \multirow{2}{*}{(W3) GEE-exch}
    & Coef & 0.0 & 0.17 (0.18) & 96.6 (95.7, 97.5) & \textbackslash & \textbackslash & \textbackslash \\
  & & & MRS & 0.0 & 0.17 (0.18) & 96.4 (95.5, 97.3) & 0.0 & 0.18 (0.19) & 95.8 (94.8, 96.8) \\
  & & \multirow{2}{*}{(W4) GEE-ind}
    & Coef & 0.0 & 0.18 (0.19) & 96.2 (95.3, 97.1) & \textbackslash & \textbackslash & \textbackslash \\
  & & & MRS & 0.0 & 0.17 (0.18) & 96.3 (95.4, 97.2) & 0.0 & 0.18 (0.19) & 96.2 (95.3, 97.1) \\
\bottomrule
\end{tabular}%
}

\end{center}
\end{table}

We examine $4$ working outcome models. These are (W1) a linear regression model based on cluster-level means described in Example \ref{example1}; (W2) a linear mixed model based on individual-level observations described in Example \ref{example2}; (W3) a marginal model fitted with GEE under the exchangeable working correlation structure; and (W4) a marginal model fitted with GEE under working independence, as described in Example \ref{example4}. For each (W1)-(W4), we consider both an unadjusted version where only the treatment indicator $A_i$ is included, and a covariate-adjusted version where additional baseline characteristics, $\{X_{1ij},X_{2ij},H_{1i},H_{2i},N_i\}$ are controlled for in the regression as linear main effects. For each one of these $8$ working models, we then obtain the model-robust standardization (MRS) estimator as well as the model-based coefficient (Coef) estimator corresponding to the treatment variable $A_i$. The standard errors for the MRS estimators are computed via cluster jackknifing, whereas those for the Coef estimators are obtained from the heteroskedasticity-robust sandwich variance estimator (for (W1) based on the \texttt{sandwich} R package), the approximate jackknife variance estimator (for (W2) based on the \texttt{clubSandwich} R package), and the Mancl and DeRouen bias-corrected variance estimator (for (W3) and (W4) based on the \texttt{geesmv} R package). These choices of standard error estimators reflect existing recommendations for finite-sample adjustments in CRTs.\cite{mancl2001covariance,leyrat2018cluster,ouyang2024maintaining} We investigate four performance metrics including the percentage bias (BIAS), Monte Carlo standard deviation (MCSD), average estimated standard error (AESE), and coverage probabilities for the 95\% confidence intervals constructed with the $t(m-1)$ distribution (\(95\%\) confidence limits of empirical coverage\cite{morris2019using} also provided to quantify uncertainty). The performance metrics are defined relative to the estimand $\Delta_C=\Delta_I=-3$.

Table \ref{tab_NICS} presents the simulation results under non-informative cluster size for a continuous outcome. We observe that the Coef estimators have negligible bias to either the cluster-ATE and individual-ATE, a result that is consistent with prior studies.\cite{wang2022two,wang2021mixed} The MRS estimators also have no larger bias than Coef estimators, regardless of the choice of regression models. Second, the variance estimator for Coef tends to underestimate the true variance, resulting in approximate or below 90\% coverage, when $m=30$ and covariate-adjusted working models (W1)-(W2) are used. In contrast, the MRS estimators, at worst, provide conservative coverage under comparable scenarios. Third, covariate adjustment can improve efficiency regardless of sample size and choice of estimation method. To summarize, under non-informative cluster size, the MRS method consistently performs no worse than the Coef method with a continuous outcome, although the latter has little bias despite model misspecification. This simulation experiment paves the way for Section \ref{s:sim;ss:continuous} where we modify the data generating mechanism to induce informative cluster size with a continuous outcome.

\subsection{Informative cluster size and a continuous outcome}\label{s:sim;ss:continuous}
We modify the data generating process in Section \ref{s:sim;ss:NICS} to induce informative cluster size. First, we simulate one binary cluster-level covariate $H_{1i}\sim \text{Bernoulli}\{\Phi(\text{sin}(N_i))\}$, and one continuous cluster-level covariate $H_{2i}\sim \mathcal{N}(2+H_{1i}N_i/10,9)$. We then generate one continuous individual-level covariate $X_{1ij}\sim\mathcal{N}(H_{1i}H_{2i}+N_i/100,16)$ and one binary individual-level covariate $X_{2ij}\sim \text{Bernoulli}\{\text{expit}(\text{log}(N_i)X_{1ij}H_{1i}+H_{2i})\}$. Unlike Section \ref{s:sim;ss:NICS}, where only the overall mean cluster size is correlated with each covariate, we now allow the actual cluster size to be correlated with each covariate. This introduces the first source of informative cluster size. The potential outcomes are simulated according to 
\begin{align*}
    Y_{ij}(a)\sim \mathcal{N}\left(\frac{H_{1i}X_{1ij}^2 }{5N_i}-\frac{N_{i}^2\text{log}(N_i)}{(E\{N_i\})^2}+\text{cos}(H_{2i})X_{2ij}+|H_{2i}|\text{sin}(X_{2ij})+\frac{N_{i}^2\text{log}(N_i)}{(E\{N_i\})^2}a+\gamma_i a,1\right),~~~~a\in\{0,1\}.
\end{align*}
where $\gamma_i\sim\mathcal{N}(0,0.2)$. Unlike Section \ref{s:sim;ss:NICS}, here the true treatment effect in each cluster is proportional to $N_i^2\log(N_i)$. As a result, the true values of cluster-level ATE and individual-level ATE depend on the distribution of $N_i$. The cluster size becomes an effect modifier, representing the second source of informative cluster size within this data generating process. The true values of cluster-ATE and individual-ATE are computed from a simulated super population of $10^7$ clusters. That is, we assign all $10^7$ clusters to both the intervention and control conditions, obtain their true potential outcomes, and calculate the target estimand based on the potential outcome contrasts. The true estimands are $\Delta_C=5.92$ versus $\Delta_I=8.15$ when $m=30$, and $\Delta_C=4.48$ versus $\Delta_I=6.25$ when $m=100$. The two sets of estimands differ because the cluster size distribution is set to differ depending on the number of clusters.

\begin{table}[ht!]
\caption{\label{tab_continuous_se_m100_cate} Simulation results under a data generating process with \textbf{informative} cluster size and a continuous outcome. LM: linear model based on cluster-level means; LMM: linear mixed model based on individual-level data; GEE-exch: marginal model fitted with GEE and an exchangeable working correlation structure; GEE-ind: marginal model fitted with GEE and working independence. Each empirical coverage probability is reported together with its Monte Carlo 95\% confidence limits (95\% CL). Additional Monte Carlo standard errors for performance metric estimates are presented in Web Table 2.}
\begin{center}
\resizebox{\textwidth}{!}{%
\begin{tabular}{cclcrrrrrr}
\toprule
\multirow{2}{*}{$m$} & Covariate & \multirow{2}{*}{Outcome model} & \multirow{2}{*}{Method}
& \multirow{2}{*}{BIAS} & MCSD & COV
& \multirow{2}{*}{BIAS} & MCSD & COV \\
& adjustment & & & & (AESE) &(95\% CL) & & (AESE) & (95\% CL)\\ \midrule
&&&& \multicolumn{3}{c}{$\Delta_C=5.92$} & \multicolumn{3}{c}{$\Delta_I=8.15$} \\
\multirow{16}{*}{$30$}
  & \multirow{8}{*}{$\times$}
      & \multirow{2}{*}{(W1) LM}
    & Coef & -0.3 & 2.16 (2.12) & 93.6 (92.2, 95.0) & -27.6 & 2.16 (2.12) & 78.5 (76.0, 81.0) \\
  & & & MRS & -0.3 & 2.16 (2.25) & 94.5 (93.2, 95.8) & -2.1 & 2.77 (2.88) & 93.7 (92.3, 95.1) \\
  & & \multirow{2}{*}{(W2) LMM}
    & Coef & -0.2 & 2.16 (2.12) & 93.6 (92.2, 95.0) & -27.6 & 2.16 (2.12) & 78.5 (76.0, 81.0) \\
  & & & MRS & -0.3 & 2.16 (2.25) & 94.5 (93.2, 95.8) & -2.1 & 2.77 (2.88) & 93.7 (92.3, 95.1) \\
  & & \multirow{2}{*}{(W3) GEE-exch}
    & Coef & -0.3 & 2.16 (2.28) & 94.6 (93.3, 95.9) & -27.6 & 2.16 (2.28) & 81.7 (79.3, 84.1) \\
  & & & MRS & -0.3 & 2.16 (2.25) & 94.5 (93.2, 95.8) & -2.1 & 2.77 (2.88) & 93.7 (92.3, 95.1) \\
  & & \multirow{2}{*}{(W4) GEE-ind}
    & Coef & 35.0 & 2.80 (2.99) & 92.7 (91.3, 94.1) & -2.0 & 2.80 (2.99) & 93.8 (92.5, 95.1) \\
  & & & MRS & -0.3 & 2.18 (2.31) & 94.5 (93.2, 95.8) & -2.0 & 2.80 (2.94) & 93.4 (92.0, 94.8) \\
[0.4ex]
  & \multirow{8}{*}{$\checkmark$}
      & \multirow{2}{*}{(W1) LM}
    & Coef & 0.1 & 1.18 (1.04) & 90.7 (88.9, 92.5) & -27.3 & 1.18 (1.04) & 47.9 (44.8, 51.0) \\
  & & & MRS & 0.1 & 1.18 (1.55) & 97.4 (96.3, 98.5) & -7.3 & 1.28 (1.71) & 94.5 (93.2, 95.8) \\
  & & \multirow{2}{*}{(W2) LMM}
    & Coef & 0.2 & 1.18 (1.04) & 90.7 (88.9, 92.5) & -27.3 & 1.18 (1.04) & 48.0 (44.9, 51.1) \\
  & & & MRS & 0.1 & 1.18 (1.43) & 96.5 (95.4, 97.6) & -7.3 & 1.28 (1.58) & 93.6 (92.3, 94.9) \\
  & & \multirow{2}{*}{(W3) GEE-exch}
    & Coef & 0.1 & 1.18 (1.46) & 96.8 (95.8, 97.8) & -27.3 & 1.18 (1.46) & 68.7 (66.0, 71.4) \\
  & & & MRS & 0.1 & 1.18 (1.43) & 96.5 (95.4, 97.6) & -7.3 & 1.28 (1.58) & 93.6 (92.3, 94.9) \\
  & & \multirow{2}{*}{(W4) GEE-ind}
    & Coef & 29.9 & 1.31 (1.66) & 83.8 (81.6, 86.0) & -5.7 & 1.31 (1.66) & 94.7 (93.5, 95.9) \\
  & & & MRS & 1.6 & 1.20 (1.49) & 97.0 (96.0, 98.0) & -5.7 & 1.31 (1.63) & 94.1 (92.8, 95.4) \\
[0.6ex]

&&&& \multicolumn{3}{c}{$\Delta_C=4.48$} & \multicolumn{3}{c}{$\Delta_I=6.25$} \\
\multirow{16}{*}{$100$}
  & \multirow{8}{*}{$\times$}
      & \multirow{2}{*}{(W1) LM}
    & Coef & -0.2 & 0.72 (0.74) & 95.9 (94.8, 97.0) & -28.4 & 0.72 (0.74) & 34.6 (31.6, 37.6) \\
  & & & MRS & -0.2 & 0.72 (0.75) & 96.3 (95.3, 97.3) & -1.1 & 0.85 (0.89) & 95.2 (94.0, 96.4) \\
  & & \multirow{2}{*}{(W2) LMM}
    & Coef & 0.0 & 0.72 (0.74) & 95.8 (94.7, 96.9) & -28.3 & 0.72 (0.74) & 35.3 (32.3, 38.3) \\
  & & & MRS & -0.2 & 0.72 (0.75) & 96.3 (95.3, 97.3) & -1.1 & 0.85 (0.89) & 95.2 (94.0, 96.4) \\
  & & \multirow{2}{*}{(W3) GEE-exch}
    & Coef & -0.3 & 0.72 (0.75) & 96.3 (95.3, 97.3) & -28.5 & 0.72 (0.75) & 35.9 (32.9, 38.9) \\
  & & & MRS & -0.2 & 0.72 (0.75) & 96.3 (95.3, 97.3) & -1.1 & 0.85 (0.89) & 95.2 (94.0, 96.4) \\
  & & \multirow{2}{*}{(W4) GEE-ind}
    & Coef & 37.9 & 0.83 (0.87) & 51.2 (48.1, 54.3) & -1.0 & 0.83 (0.87) & 95.5 (94.4, 96.6) \\
  & & & MRS & -0.2 & 0.74 (0.76) & 96.2 (95.2, 97.2) & -1.0 & 0.83 (0.86) & 95.4 (94.3, 96.5) \\
[0.4ex]
  & \multirow{8}{*}{$\checkmark$}
      & \multirow{2}{*}{(W1) LM}
    & Coef & 0.2 & 0.47 (0.44) & 92.9 (91.4, 94.4) & -28.1 & 0.47 (0.44) & 3.7 (2.8, 4.6) \\
  & & & MRS & 0.2 & 0.47 (0.49) & 95.0 (93.9, 96.1) & -2.1 & 0.52 (0.54) & 93.8 (92.6, 95.0) \\
  & & \multirow{2}{*}{(W2) LMM}
    & Coef & 0.7 & 0.47 (0.44) & 93.1 (91.6, 94.6) & -27.8 & 0.47 (0.44) & 3.7 (2.8, 4.6) \\
  & & & MRS & 0.2 & 0.47 (0.48) & 94.9 (93.8, 96.0) & -2.1 & 0.52 (0.52) & 93.4 (92.2, 94.6) \\
  & & \multirow{2}{*}{(W3) GEE-exch}
    & Coef & 0.3 & 0.47 (0.48) & 94.9 (93.8, 96.0) & -28.0 & 0.47 (0.48) & 4.9 (3.8, 6.0) \\
  & & & MRS & 0.2 & 0.47 (0.48) & 94.9 (93.8, 96.0) & -2.1 & 0.52 (0.52) & 93.4 (92.2, 94.6) \\
  & & \multirow{2}{*}{(W4) GEE-ind}
    & Coef & 37.1 & 0.51 (0.53) & 10.8 (9.3, 12.3) & -1.6 & 0.51 (0.53) & 93.6 (92.4, 94.8) \\
  & & & MRS & 0.6 & 0.47 (0.49) & 94.7 (93.6, 95.8) & -1.6 & 0.51 (0.52) & 93.4 (92.2, 94.6) \\
\bottomrule
\end{tabular}%
}
\end{center}
\end{table}

Table \ref{tab_continuous_se_m100_cate} presents the simulation results when the exact same 4 working models (with and without linear covariate adjustment) in Section \ref{s:sim;ss:NICS} are fitted; the presentation of Table \ref{tab_continuous_se_m100_cate} follows Table \ref{tab_NICS} to enable direct comparisons. It is apparent that, under informative cluster size, the performance of the Coef estimator critically depends on the choice of the outcome regression model. In particular, regardless of covariate adjustment, the Coef estimator from the independence GEE (W4) is biased for estimating cluster-ATE, while the Coef estimators from the remaining models, (W1)-(W3), are biased for estimating individual-ATE. This bias then becomes the primary source for under-coverage. Such biases have been pointed out from previous studies,\cite{wang2022two,su2021model} and arise because the fitted regression model does not employ the right weighting scheme that matches the marginal estimands $\Delta_C$ or $\Delta_I$ (also see Kahan et al.\cite{kahan2023demystifying} and Lee et al.\cite{lee2025should} for detailed explanation about why Coef from LMM could be biased under informative cluster size). In sharp contrast, the relative bias is negligible and the empirical coverage probabilities are close to $95\%$ for MRS estimators, regardless of the choice of outcome regression models and covariate adjustment. Finally, covariate adjustment under MRS also leads to no worse finite-sample efficiency compared to no covariate adjustment. In summary, although Coef and MRS perform relatively similarly under non-informative cluster size, the findings in Section \ref{s:sim;ss:continuous} suggest that (1) the performance of Coef highly depends on the choice of outcome model and Coef may target neither cluster-ATE or individual-ATE; (2) even with a given outcome model whose Coef estimator is biased for the estimand of interest, applying our MRS procedure can fix the bias.

For readers interested in isolating specific sources of model misspecification on the results patterns, we have considered a simpler setup where informative cluster size is introduced only through the treatment effect (rather than complex covariate dependence and ICC heterogeneity). The detailed simulation results are reported in Web Appendix Section D, and the overall trends are similar to those observed in Table \ref{tab_continuous_se_m100_cate}. This reinforces the robustness of our main findings.

\subsection{Non-informative cluster size and a binary outcome}\label{s:sim;ss:NICS_binary}

We next present a parallel set of simulations for a binary outcome with a ratio estimand. As a benchmark for evaluating the Coef and MRS methods, we first describe the simulations conducted under non-informative cluster size. Cluster-level covariates are simulated according to $H_{1i}\sim\text{Bernoulli}(0.5)$ and $H_{2i}\sim\mathcal{N}(3+H_{1i},1)$, and individual-level covariates are generated according to $X_{1ij}\sim\mathcal{N}(H_{1i}+H_{2i}/20+1,16)$ and $X_{2ij}\sim\text{Bernoulli}\{\text{expit}(4H_{1i}X_{1ij}+H_{2i})\}$. The potential outcomes are simulated from 
\begin{align*}
    Y_{ij}(a)\sim \text{Bernoulli}\left(\text{expit}\left(-0.8+\frac{X_{1ij}^2}{100}+H_{1i}+\text{cos}(H_{2i})X_{2ij}+\frac{|H_{2i}|}{5}+0.8a+\gamma_i a\right)\right),~~~~a\in\{0,1\}.
\end{align*}
We focus on the estimands on the log odds ratio scale, that is, using $f(x,y)=\text{log}\{x(1-y)y^{-1}(1-x)^{-1}\}$. We simulate a super population of $m=10^7$ clusters and obtain the cluster-ATE and individual-ATE as $0.65$. In the above data generating process, the cluster size does not affect the generation of any covariate nor the potential outcome; hence there is no informative cluster size and cluster-ATE and individual-ATE coincide on the log odds ratio scale.

\begin{table}[ht!]
\caption{\label{tab_NICS_binary} Simulation results under a data generating process with \textbf{non-informative} cluster size and a binary outcome. GLM: logistic generalized linear model based on cluster-level means; GLMM: logistic generalized linear mixed model based on individual-level data; GEE-exch: marginal model fitted with GEE and an exchangeable working correlation structure; GEE-ind: marginal model fitted with GEE and working independence. Each empirical coverage probability is reported together with its Monte Carlo 95\% confidence limits (95\% CL). Additional Monte Carlo standard errors for performance metric estimates are presented in Web Table 3.}
\begin{center}
\resizebox{\textwidth}{!}{%
\begin{tabular}{cclcrrrrrr}
\toprule
\multirow{2}{*}{$m$} & Covariate & \multirow{2}{*}{Outcome model} & \multirow{2}{*}{Method}
& \multirow{2}{*}{BIAS} & MCSD & COV
& \multirow{2}{*}{BIAS} & MCSD & COV \\
& adjustment & & & & (AESE) & (95\% CL) & & (AESE) & (95\% CL)\\ \midrule
&&&& \multicolumn{3}{c}{$\Delta_C=0.65$} & \multicolumn{3}{c}{$\Delta_I=0.65$} \\
\multirow{16}{*}{$30$}
  & \multirow{8}{*}{$\times$}
      & \multirow{2}{*}{(W1) LM}
    & Coef & 0.9 & 0.32 (0.32) & 95.5 (94.5, 96.5) & \textbackslash & \textbackslash & \textbackslash \\
  & & & MRS & 0.9 & 0.32 (0.32) & 95.1 (94.1, 96.1) & 2.3 & 0.35 (0.35) & 95.0 (94.0, 96.0) \\
  & & \multirow{2}{*}{(W2) LMM}
    & Coef & 19.2 & 0.37 (0.35) & 92.8 (91.5, 94.1) & \textbackslash & \textbackslash & \textbackslash \\
  & & & MRS & 0.9 & 0.32 (0.32) & 95.1 (94.1, 96.1) & 2.3 & 0.35 (0.35) & 94.7 (93.7, 95.7) \\
  & & \multirow{2}{*}{(W3) GEE-exch}
    & Coef & 1.1 & 0.32 (0.33) & 95.7 (94.7, 96.7) & \textbackslash & \textbackslash & \textbackslash \\
  & & & MRS & 0.9 & 0.32 (0.32) & 95.2 (94.2, 96.2) & 2.3 & 0.35 (0.35) & 94.9 (93.9, 95.9) \\
  & & \multirow{2}{*}{(W4) GEE-ind}
    & Coef & 2.1 & 0.35 (0.36) & 94.5 (93.3, 95.7) & \textbackslash & \textbackslash & \textbackslash \\
  & & & MRS & 0.7 & 0.32 (0.33) & 95.4 (94.4, 96.4) & 2.1 & 0.35 (0.36) & 94.2 (93.1, 95.3) \\
[0.4ex]
  & \multirow{8}{*}{$\checkmark$}
      & \multirow{2}{*}{(W1) LM}
    & Coef & 16.8 & 0.22 (0.20) & 89.7 (88.1, 91.3) & \textbackslash & \textbackslash & \textbackslash \\
  & & & MRS & 0.7 & 0.19 (0.24) & 96.2 (95.2, 97.2) & 1.0 & 0.19 (0.24) & 96.5 (95.6, 97.4) \\
  & & \multirow{2}{*}{(W2) LMM}
    & Coef & 19.3 & 0.21 (0.18) & 84.9 (83.1, 86.7) & \textbackslash & \textbackslash & \textbackslash \\
  & & & MRS & 0.5 & 0.18 (0.21) & 95.3 (94.3, 96.3) & 0.8 & 0.18 (0.21) & 95.9 (95.0, 96.8) \\
  & & \multirow{2}{*}{(W3) GEE-exch}
    & Coef & 16.2 & 0.21 (0.24) & 93.6 (92.4, 94.8) & \textbackslash & \textbackslash & \textbackslash \\
  & & & MRS & 0.5 & 0.18 (0.22) & 95.2 (94.2, 96.2) & 0.8 & 0.18 (0.22) & 95.9 (95.0, 96.8) \\
  & & \multirow{2}{*}{(W4) GEE-ind}
    & Coef & 16.5 & 0.21 (0.25) & 94.0 (92.9, 95.1) & \textbackslash & \textbackslash & \textbackslash \\
  & & & MRS & 0.5 & 0.18 (0.21) & 95.5 (94.6, 96.4) & 0.9 & 0.18 (0.22) & 95.6 (94.6, 96.6) \\
[0.6ex]

&&&& \multicolumn{3}{c}{$\Delta_C=0.65$} & \multicolumn{3}{c}{$\Delta_I=0.65$} \\
\multirow{16}{*}{$100$}
  & \multirow{8}{*}{$\times$}
      & \multirow{2}{*}{(W1) LM}
    & Coef & -6.0 & 0.19 (0.18) & 92.2 (91.0, 93.4) & \textbackslash & \textbackslash & \textbackslash \\
  & & & MRS & -6.0 & 0.19 (0.18) & 92.3 (91.1, 93.5) & -3.5 & 0.20 (0.19) & 92.7 (91.5, 93.9) \\
  & & \multirow{2}{*}{(W2) LMM}
    & Coef & 10.2 & 0.22 (0.20) & 92.4 (91.2, 93.6) & \textbackslash & \textbackslash & \textbackslash \\
  & & & MRS & -5.8 & 0.19 (0.18) & 92.1 (90.9, 93.3) & -3.4 & 0.20 (0.19) & 92.7 (91.5, 93.9) \\
  & & \multirow{2}{*}{(W3) GEE-exch}
    & Coef & -5.1 & 0.19 (0.18) & 92.6 (91.4, 93.8) & \textbackslash & \textbackslash & \textbackslash \\
  & & & MRS & -5.9 & 0.19 (0.18) & 92.3 (91.1, 93.5) & -3.5 & 0.20 (0.19) & 92.8 (91.6, 94.0) \\
  & & \multirow{2}{*}{(W4) GEE-ind}
    & Coef & -3.4 & 0.20 (0.19) & 93.2 (92.0, 94.4) & \textbackslash & \textbackslash & \textbackslash \\
  & & & MRS & -5.8 & 0.19 (0.18) & 92.1 (90.9, 93.3) & -3.4 & 0.20 (0.19) & 92.7 (91.5, 93.9) \\
[0.4ex]
  & \multirow{8}{*}{$\checkmark$}
      & \multirow{2}{*}{(W1) LM}
    & Coef & 11.1 & 0.13 (0.13) & 90.7 (89.2, 92.2) & \textbackslash & \textbackslash & \textbackslash \\
  & & & MRS & -3.3 & 0.12 (0.12) & 96.1 (95.2, 97.0) & -1.9 & 0.11 (0.12) & 96.4 (95.5, 97.3) \\
  & & \multirow{2}{*}{(W2) LMM}
    & Coef & 14.8 & 0.13 (0.13) & 87.4 (85.7, 89.1) & \textbackslash & \textbackslash & \textbackslash \\
  & & & MRS & -3.1 & 0.11 (0.12) & 95.0 (94.0, 96.0) & -1.8 & 0.11 (0.12) & 96.0 (95.1, 96.9) \\
  & & \multirow{2}{*}{(W3) GEE-exch}
    & Coef & 11.4 & 0.12 (0.13) & 92.5 (91.2, 93.8) & \textbackslash & \textbackslash & \textbackslash \\
  & & & MRS & -3.2 & 0.11 (0.12) & 95.3 (94.4, 96.2) & -1.9 & 0.11 (0.12) & 95.8 (94.9, 96.7) \\
  & & \multirow{2}{*}{(W4) GEE-ind}
    & Coef & 12.0 & 0.13 (0.14) & 92.5 (91.2, 93.8) & \textbackslash & \textbackslash & \textbackslash \\
  & & & MRS & -3.2 & 0.12 (0.12) & 95.3 (94.4, 96.2) & -1.9 & 0.11 (0.12) & 95.7 (94.8, 96.6) \\
\bottomrule
\end{tabular}%
}

\end{center}
\end{table}

For each simulated dataset, we fit four working outcome models including: (W5) a logistic regression model based on cluster-level proportions described in Example \ref{example1}; (W6) a logistic generalized linear mixed model based on individual-level observations described in Example \ref{example3}; (W7) a marginal model fitted with GEE under the \verb"logit" link function and an exchangeable working correlation structure; and (W8) a marginal model fitted with GEE under the \verb"logit" link function and working independence, as described in Example \ref{example4}. We consider both an unadjusted estimator and a covariate-adjusted estimator (linear adjustment) for each working outcome model, and compare the Coef and MRS estimators. The standard errors for the Coef estimators are calculated using the model-based variance estimator (for (W5) based on the \texttt{glm} R package, for (W6) based on the \verb"lme4" R package, as is common practice) and the Mancl and DeRouen bias-corrected variance estimator (for (W7) and (W8) based on the \verb"geesmv" R package).

Table \ref{tab_NICS_binary} summarizes the results under non-informative cluster size for a binary outcome. Without covariate adjustment, the Coef estimators from GLM (W5) and GEE (W6-W7) provide valid inference. This is expected because (i) the treatment coefficient from the logistic regression based on cluster-level proportions automatically targets the cluster-ATE (which also equals to individual-ATE without informative cluster size);\cite{kahan2023demystifying} (ii) and the GEE treatment coefficient without additional covariates converges to both cluster-ATE and individual-ATE, regardless of the working correlation structure. However, due to non-collapsibility, the Coef estimator from a logistic GLMM based on individual observations remains biased even without covariate adjustment. Furthermore, different from Table \ref{tab_NICS}, the Coef estimators adjusting for covariates are generally biased for the cluster-ATE and individual-ATE (leading to under-coverage), regardless of outcome models. This is because, when a covariate-adjusted logistic GLM, GLMM or GEE misspecifies the functional form of covariates or the true random-effects structure, their Coef estimator may converge to an ambiguous probability limit. Different from a difference estimand, when a ratio estimand is considered, the performance of the Coef estimator depends on the choice of outcome model and covariate adjustment, even without informative cluster size. In contrast, the MRS estimators demonstrate negligible bias and achieve nominal coverage regardless of the choice of outcome model and covariate adjustment. This simulation experiment paves the way for Section \ref{s:sim;ss:binary} under informative cluster size.

\subsection{Informative cluster size and a binary outcome}\label{s:sim;ss:binary}

To build in informative cluster size, we modify the data generating process from Section \ref{s:sim;ss:NICS_binary} as follows. The continuous cluster-level covariate is simulated from $H_{2i}\sim \mathcal{N}(2+H_{1i}+N_i/E\{N_i\},1)$; the continuous individual-level covariate is drawn from $X_{1ij}\sim\mathcal{N}(H_{1i}+H_{2i}/20+N_i/100,16)$, \(X_{2ij} \sim \text{Bernoulli}\{\text{expit}(\log(N_{ij})H_{1i}X_{1ij} + H_{2i} )\} \); and then the binary potential outcomes are generated from 
\begin{align*}
    Y_{ij}(a)&\sim \text{Bernoulli}\left(\text{expit}\left(-\frac{N_{i}^2\text{log}(N_i)}{5(E\{N_i\})^2}+\frac{X_{1ij}^2}{2N_i}+H_{1i}+\text{cos}(H_{2i})X_{2ij}+\frac{|H_{2i}|}{5}+\frac{N_{i}^2\text{log}(N_i)}{5(E\{N_i\})^2}a+\gamma_i a\right)\right),~~~~a\in\{0,1\}.
\end{align*}
Similar to the data generating process in Secion \ref{s:sim;ss:continuous}, the above data generating process induces informative cluster size by (1) associating the covariates with cluster size (such that $N_i$ is marginally correlated with potential outcome), and (2) including an interaction between cluster size and treatment in the potential outcome model (such that $N_i$ is an effect modifier). 
The same set of outcome models used in Section \ref{s:sim;ss:NICS_binary} are used in this simulation experiment.

Table \ref{tab_binary_se_m100_cate} presents the simulation results under informative cluster size with a binary outcome. The MRS estimators again present negligible bias and nominal coverage for both the cluster-ATE and the individual-ATE, regardless of outcome model specification. Similar to the continuous outcome setting, covariate adjustment can improve precision by reducing the MCSD metric under MRS; however, the improvement in efficiency with a binary outcome appears more modest (but still non-trivial) than with a continuous outcome. The Coef estimators, however, are generally not guaranteed to target cluster-ATE or individual-ATE, and hence the bias and coverage can be sensitive to model choice and covariate adjustment. The only two exceptions are that the GLM based on cluster-level proportions without covariates (previously proved to be consistent for the cluster-ATE\cite{kahan2023demystifying}), and the independence GEE without covariates (previously proved to be consistent for the individual-ATE\cite{wang2022two,kahan2023demystifying}). Surprisingly, we find through this simulation that the covariate-adjusted logistic GLMM has small bias for estimating the individual-ATE (but has substantial bias for the cluster-ATE). However, this is an empirical observation that may well be specific to the current simulation setting and we are unable to analytically justify this observation across all data generating mechanisms. In summary, the simulations with a binary outcome continues to endorse the patterns observed under a continuous outcome. That is, the performance of Coef depends on the choice of outcome model. But for any given outcome model, applying our MRS procedure to the model output can help target cluster-ATE and individual-ATE with negligible bias.

\begin{table}[ht!]
\caption{\label{tab_binary_se_m100_cate} Simulation results under a data generating process with \textbf{informative} cluster size and a binary outcome. GLM: logistic generalized linear model based on cluster-level means; GLMM: logistic generalized linear mixed model based on individual-level data; GEE-exch: marginal model fitted with GEE and an exchangeable working correlation structure; GEE-ind: marginal model fitted with GEE and working independence. Each empirical coverage probability is reported together with its Monte Carlo 95\% confidence limits (95\% CL). Additional Monte Carlo standard errors for performance metric estimates are presented in Web Table 4.}
\begin{center}
\resizebox{\textwidth}{!}{%
\begin{tabular}{cclcrrrrrr}
\toprule
\multirow{2}{*}{$m$} & Covariate & \multirow{2}{*}{Outcome model} & \multirow{2}{*}{Method}
& \multirow{2}{*}{BIAS} & MCSD & COV
& \multirow{2}{*}{BIAS} & MCSD & COV \\
& adjustment & & & & (AESE) & (95\% CL) & & (AESE) & (95\% CL) \\ \midrule
&&&& \multicolumn{3}{c}{$\Delta_C=0.91$} & \multicolumn{3}{c}{$\Delta_I=1.24$} \\
\multirow{16}{*}{$30$}
  & \multirow{8}{*}{$\times$}
      & \multirow{2}{*}{(W1) LM}
    & Coef & 1.3 & 0.36 (0.36) & 95.5 (94.5, 96.5) & -25.9 & 0.36 (0.36) & 85.2 (83.3, 87.1) \\
  & & & MRS & 1.3 & 0.36 (0.36) & 95.6 (94.6, 96.6) & 0.3 & 0.41 (0.42) & 95.4 (94.4, 96.4) \\
  & & \multirow{2}{*}{(W2) LMM}
    & Coef & 31.6 & 0.45 (0.43) & 90.0 (88.4, 91.6) & -3.7 & 0.45 (0.43) & 93.1 (91.9, 94.3) \\
  & & & MRS & 1.3 & 0.36 (0.37) & 95.8 (94.8, 96.8) & 0.1 & 0.41 (0.42) & 95.3 (94.3, 96.3) \\
  & & \multirow{2}{*}{(W3) GEE-exch}
    & Coef & 3.2 & 0.36 (0.37) & 95.5 (94.5, 96.5) & -24.5 & 0.36 (0.37) & 86.9 (85.1, 88.7) \\
  & & & MRS & 1.3 & 0.36 (0.36) & 95.7 (94.7, 96.7) & 0.3 & 0.41 (0.42) & 95.4 (94.4, 96.4) \\
  & & \multirow{2}{*}{(W4) GEE-ind}
    & Coef & 37.0 & 0.40 (0.42) & 89.8 (88.2, 91.4) & 0.3 & 0.40 (0.42) & 95.6 (94.6, 96.6) \\
  & & & MRS & 1.3 & 0.37 (0.37) & 95.4 (94.4, 96.4) & 0.3 & 0.40 (0.42) & 95.5 (94.5, 96.5) \\
[0.4ex]
  & \multirow{8}{*}{$\checkmark$}
      & \multirow{2}{*}{(W1) LM}
    & Coef & 25.9 & 0.32 (0.29) & 86.0 (84.2, 87.8) & -7.9 & 0.32 (0.29) & 89.7 (88.1, 91.3) \\
  & & & MRS & 3.6 & 0.25 (0.34) & 97.0 (96.1, 97.9) & -4.1 & 0.26 (0.36) & 97.8 (97.0, 98.6) \\
  & & \multirow{2}{*}{(W2) LMM}
    & Coef & 39.6 & 0.32 (0.27) & 72.0 (69.6, 74.4) & 2.1 & 0.32 (0.27) & 89.2 (87.6, 90.8) \\
  & & & MRS & 3.6 & 0.25 (0.30) & 96.3 (95.4, 97.2) & -4.3 & 0.26 (0.31) & 96.6 (95.7, 97.5) \\
  & & \multirow{2}{*}{(W3) GEE-exch}
    & Coef & 27.9 & 0.30 (0.36) & 90.7 (89.2, 92.2) & -6.4 & 0.30 (0.36) & 96.0 (95.1, 96.9) \\
  & & & MRS & 3.3 & 0.25 (0.29) & 96.1 (95.2, 97.0) & -4.4 & 0.26 (0.31) & 96.7 (95.8, 97.6) \\
  & & \multirow{2}{*}{(W4) GEE-ind}
    & Coef & 58.4 & 0.31 (0.39) & 75.2 (72.9, 77.5) & 15.9 & 0.31 (0.39) & 94.2 (93.2, 95.2) \\
  & & & MRS & 4.3 & 0.25 (0.30) & 96.3 (95.4, 97.2) & -3.2 & 0.26 (0.32) & 96.2 (95.3, 97.1) \\
[0.6ex]

&&&& \multicolumn{3}{c}{$\Delta_C=0.71$} & \multicolumn{3}{c}{$\Delta_I=0.97$} \\
\multirow{16}{*}{$100$}
  & \multirow{8}{*}{$\times$}
      & \multirow{2}{*}{(W1) LM}
    & Coef & -0.8 & 0.19 (0.19) & 96.4 (95.5, 97.3) & -27.4 & 0.19 (0.19) & 72.8 (70.4, 75.2) \\
  & & & MRS & -0.8 & 0.19 (0.19) & 96.0 (95.0, 97.0) & -0.9 & 0.22 (0.22) & 95.9 (94.9, 96.9) \\
  & & \multirow{2}{*}{(W2) LMM}
    & Coef & 28.3 & 0.23 (0.23) & 85.5 (83.7, 87.3) & -6.1 & 0.23 (0.23) & 94.5 (93.4, 95.6) \\
  & & & MRS & -0.8 & 0.19 (0.19) & 96.0 (95.0, 97.0) & -0.9 & 0.22 (0.22) & 95.8 (94.8, 96.8) \\
  & & \multirow{2}{*}{(W3) GEE-exch}
    & Coef & 5.1 & 0.19 (0.19) & 95.7 (94.8, 96.6) & -23.1 & 0.19 (0.19) & 79.6 (77.5, 81.7) \\
  & & & MRS & -0.9 & 0.19 (0.19) & 96.0 (95.0, 97.0) & -0.9 & 0.22 (0.22) & 96.0 (95.1, 96.9) \\
  & & \multirow{2}{*}{(W4) GEE-ind}
    & Coef & 35.1 & 0.21 (0.22) & 77.7 (75.5, 79.9) & -1.1 & 0.21 (0.22) & 96.0 (95.1, 96.9) \\
  & & & MRS & -1.1 & 0.19 (0.19) & 96.3 (95.4, 97.2) & -1.1 & 0.21 (0.21) & 96.0 (95.1, 96.9) \\
[0.4ex]
  & \multirow{8}{*}{$\checkmark$}
      & \multirow{2}{*}{(W1) LM}
    & Coef & 20.8 & 0.16 (0.15) & 81.2 (79.1, 83.3) & -11.6 & 0.16 (0.15) & 85.2 (83.3, 87.1) \\
  & & & MRS & 1.5 & 0.13 (0.14) & 96.6 (95.8, 97.4) & -1.1 & 0.13 (0.15) & 95.4 (94.5, 96.3) \\
  & & \multirow{2}{*}{(W2) LMM}
    & Coef & 41.4 & 0.16 (0.15) & 49.2 (46.5, 51.9) & 3.5 & 0.16 (0.15) & 93.4 (92.2, 94.6) \\
  & & & MRS & 1.4 & 0.13 (0.14) & 96.2 (95.4, 97.0) & -1.2 & 0.13 (0.14) & 94.9 (94.0, 95.8) \\
  & & \multirow{2}{*}{(W3) GEE-exch}
    & Coef & 32.8 & 0.15 (0.16) & 69.7 (67.4, 72.0) & -2.8 & 0.15 (0.16) & 95.6 (94.8, 96.4) \\
  & & & MRS & 1.4 & 0.13 (0.14) & 96.0 (95.2, 96.8) & -1.1 & 0.13 (0.14) & 94.8 (93.9, 95.7) \\
  & & \multirow{2}{*}{(W4) GEE-ind}
    & Coef & 57.6 & 0.15 (0.16) & 26.2 (23.8, 28.6) & 15.4 & 0.15 (0.16) & 86.2 (84.3, 88.1) \\
  & & & MRS & 1.7 & 0.13 (0.14) & 96.1 (95.3, 96.9) & -0.8 & 0.13 (0.14) & 94.7 (93.8, 95.6) \\
\bottomrule
\end{tabular}%
}
\end{center}
\end{table}

\subsection{Additional simulations with a smaller sample size and under constrained randomization}\label{sec:others}
We also replicate the above simulation design under informative cluster size (I) when the number of clusters is small and (II) when the study adopts covariate-constrained randomization. The additional simulations under (I) are used to explore whether MRS remains robust to estimate cluster-ATE and individual-ATE when $m$ is as small as $12$. The additional simulations under (II) are used to explore whether MRS retains its favorable performance characteristics over Coef under a more complex randomization scheme. Detailed configurations are provided in {Web Appendix E}, with full numerical results summarized in {Web Tables 6-11}. Overall, when $m\leq20$, MRS remains robust to the choice of outcome model, but the jackknife approach can sometimes yield conservative inference. 
Next, under covariate-constrained randomization, the MRS estimators continue to exhibit negligible bias and maintain nominal or slightly conservative coverage. In contrast, the performance of the Coef estimators remains sensitive to the choice of working outcome model, and constrained randomization does not mitigate the potential bias inherent to certain Coef estimators under informative cluster size.

\subsection{Testing for informative cluster size}\label{sec:ics_test}

Finally, we conduct a simulation study to assess the operating characteristics of the test for detecting informative cluster size, with both continuous and binary outcomes. We consider \(m = 30\) and \(m = 100\) clusters, with expected cluster size \(mE(N_i) = 3000\), where cluster sizes \(N_i\) are independently drawn from a uniform distribution on \([20, 180]\) for \(m=30\), and on \([6, 54]\) for \(m=100\). For continuous outcomes, the cluster-level and individual-level covariates are generated according to Section \ref{s:sim;ss:NICS}, and the true potential outcome follows
\[
Y_{ij}(a) \sim \mathcal{N}\left( \frac{H_{1i} X_{1ij}^2}{5{E}[N_i]} + \cos(H_{2i}) X_{2ij} + |H_{2i}| \sin(X_{2ij}) + \left\{ \frac{\delta N_i^2 \log(N_i)}{({E}[N_i])^2} + 1 \right\} a + \gamma_i a, \, 1 \right),~~~~a\in\{0,1\}.
\]
Here, the parameter \(\delta\) controls the degree of informative cluster size; when $\delta=0$, the cluster size is no longer an effect modifier and the data generating process degenerates to that in Section \ref{s:sim;ss:NICS}. Following Section \ref{sec:ICS_test}, we test \(H_0: \Delta_C = \Delta_I\) using the test statistic $\widehat{D} = \widehat{\Delta}_C - \widehat{\Delta}_I =(\widehat{\mu}_C(1)-\widehat{\mu}_C(0))-(\widehat{\mu}_I(1)-\widehat{\mu}_I(0))$. The same outcome models (W1)–(W4) with linear covariate adjustment from Section \ref{s:sim;ss:NICS} are applied, and the average rejection rates across 1000 simulations (based on 0.05 level tests) are computed.

For binary outcomes, the data generation mechanism is adapted from Section \ref{s:sim;ss:NICS_binary}, by modifying the potential outcome simulations with
\[
Y_{ij}(a) \sim \text{Bernoulli}\left(\text{expit}\left(\frac{X_{1ij}^2}{2{E}[N_i]} + \frac{H_{1i}}{2} + \cos(H_{2i})X_{2ij} + \frac{|H_{2i}|}{10} + \left\{ \frac{\delta N_i^2 \log(N_i/E[N_i])}{5({E}[N_i])^2} + 1 \right\} a + \gamma_i a \right)\right),~~~~a\in\{0,1\}.
\]
Similarly, $\delta=0$ stands for non-informative cluster size, and a larger value of $\delta$ indicates higher degree of informative cluster size. The null hypothesis \(H_0: \Delta_C = \Delta_I\) is tested based on the test statistic $\widehat{D} = \text{logit}(\widehat{\Delta}_C) - \text{logit}(\widehat{\Delta}_I) =\left[\log\{\widehat{\mu}_C(1)/(1-\widehat{\mu}_C(1))\}-\log\{\widehat{\mu}_C(0)/(1-\widehat{\mu}_C(0))\}\right]-\left[\log\{\widehat{\mu}_I(1)/(1-\widehat{\mu}_I(1))\}-\log\{\widehat{\mu}_I(0)/(1-\widehat{\mu}_I(0))\}\right]$. The same outcome models (W5)–(W8) from Section \ref{s:sim;ss:NICS_binary} with linear covariate adjustment are used and the average rejection rates across 1000 simulations are recorded.

Table \ref{tab:ics_test} summarizes the empirical rejection rates for the proposed informative cluster size test across different outcome types, sample sizes, effect sizes (\(\delta\)), under different outcome models. The empirical rejection rates refer to the empirical type I error rates when $\delta=0$ and to empirical power otherwise. With $m=30$, the test appears conservative across outcome models for both continuous and binary outcomes. As \(\delta\) increases, the cluster size becomes more informative, leading to steadily increasing test power. With \(m=100\), the trends are similar, but type I error rates remain close to the nominal level. For continuous outcomes, using cluster-level methods and (generalized) linear mixed models appear to provide similar but slightly higher test power than using GEE. For binary outcomes, the cluster-level method leads to a more powerful test, especially when $m=30$. 

\begin{table}[ht]
\centering
\caption{Empirical rejection rates (\%) for the informative cluster size tests across different outcome types, sample sizes, degree of informative cluster size, and outcome models. The true values $\Delta_C$ and $\Delta_I$ are on the difference scale and log risk ratio scale for continuous outcome and binary outcome, respectively.}
\label{tab:ics_test}
\begin{tabular}{lllllcccc}
\toprule
\textbf{Outcome type} & \textbf{$m$} & \(\boldsymbol{\delta}\) & \(\Delta_C\)  & \(\Delta_I\) & \textbf{LM} & \textbf{LMM} & \textbf{GEE-exch} & \textbf{GEE-ind} \\
\midrule
\multirow{10}{*}{Continuous} 
  & \multirow{5}{*}{30}  
    & 0 & 1.000 & 1.000 & 1.6 & 1.6 & 1.1 & 1.1 \\
  &   & 0.2 & 2.183 &2.630  & 49.9 & 49.8 & 44.0 & 44.0 \\
  &   &0.6 & 4.551 & 5.893 & 92.7 & 92.7 & 87.0 & 87.0 \\
  &   &1.0 & 6.921 & 9.161 & 94.8 & 94.8 & 90.5 & 90.5 \\
  &   & 2.4 & 15.205 & 20.572 & 95.0 & 95.0 & 91.8 & 91.8 \\
\cmidrule{2-9}
  & \multirow{5}{*}{100} 
    & 0.00 &1.000 &1.000 &  5.6 & 5.6 & 4.9 & 4.9 \\
  &   & 0.05 &1.224 & 1.313 & 17.6 & 17.6 & 16.8 & 16.8 \\
  &   & 0.10 &1.448 &1.624 & 51.2 & 51.3 & 49.7 & 49.7 \\
  &   & 0.15 &1.672 &1.936 & 84.0 & 83.7 & 81.2 & 81.2 \\
  &   & 0.20 &1.895 & 2.248& 97.1 & 97.1 & 96.1 & 96.1 \\
\midrule
\textbf{Outcome type} & \textbf{$m$} &  \(\boldsymbol{\delta}\) & \(\Delta_C\) & \(\Delta_I\) &  \textbf{GLM} & \textbf{GLMM} & \textbf{GEE-exch} & \textbf{GEE-ind} \\
\midrule 
\multirow{10}{*}{Binary} 
  & \multirow{5}{*}{30}  
    & 0.00 & 0.878 & 0.878 & 1.5 & 1.7 & 1.7 & 1.0 \\
  &   & 10.00 & 1.161 & 1.491 &  52.6 & 46.8 & 51.3 & 46.2 \\
  &   & 30.00 & 1.016 & 1.554 & 82.6 & 71.8  & 78.9 & 75.4 \\
  &   & 50.00 & 0.756 & 1.375 & 88.1 & 80.8 & 83.9  & 81.3 \\
  &   & 70.00 & 0.525 & 1.201 & 89.9 & 86.0 & 86.9 & 84.9 \\
\cmidrule{2-9}
  & \multirow{5}{*}{100} 
    & 0.00 & 0.863 & 0.863 & 3.8 & 4.0 & 4.1 &  3.6 \\
  &   & 2.00 & 0.968 & 1.057 & 20.3 & 19.3 & 19.8 & 19.4 \\
  &   & 4.00 & 1.045 & 1.214 & 58.8 & 58.9 & 59.0 & 58.6 \\
  &   & 6.00 & 1.095 & 1.329 & 84.3 & 83.5 & 84.7 & 83.1 \\
  &   & 8.00 & 1.125 & 1.415 & 94.4 & 94.6 & 94.5 & 93.8 \\
\bottomrule
\end{tabular}
\end{table}

\section{An illustrative data example}\label{sec:data}


\begin{figure}[ht!]
    \centering
    \includegraphics[width=0.8\linewidth]{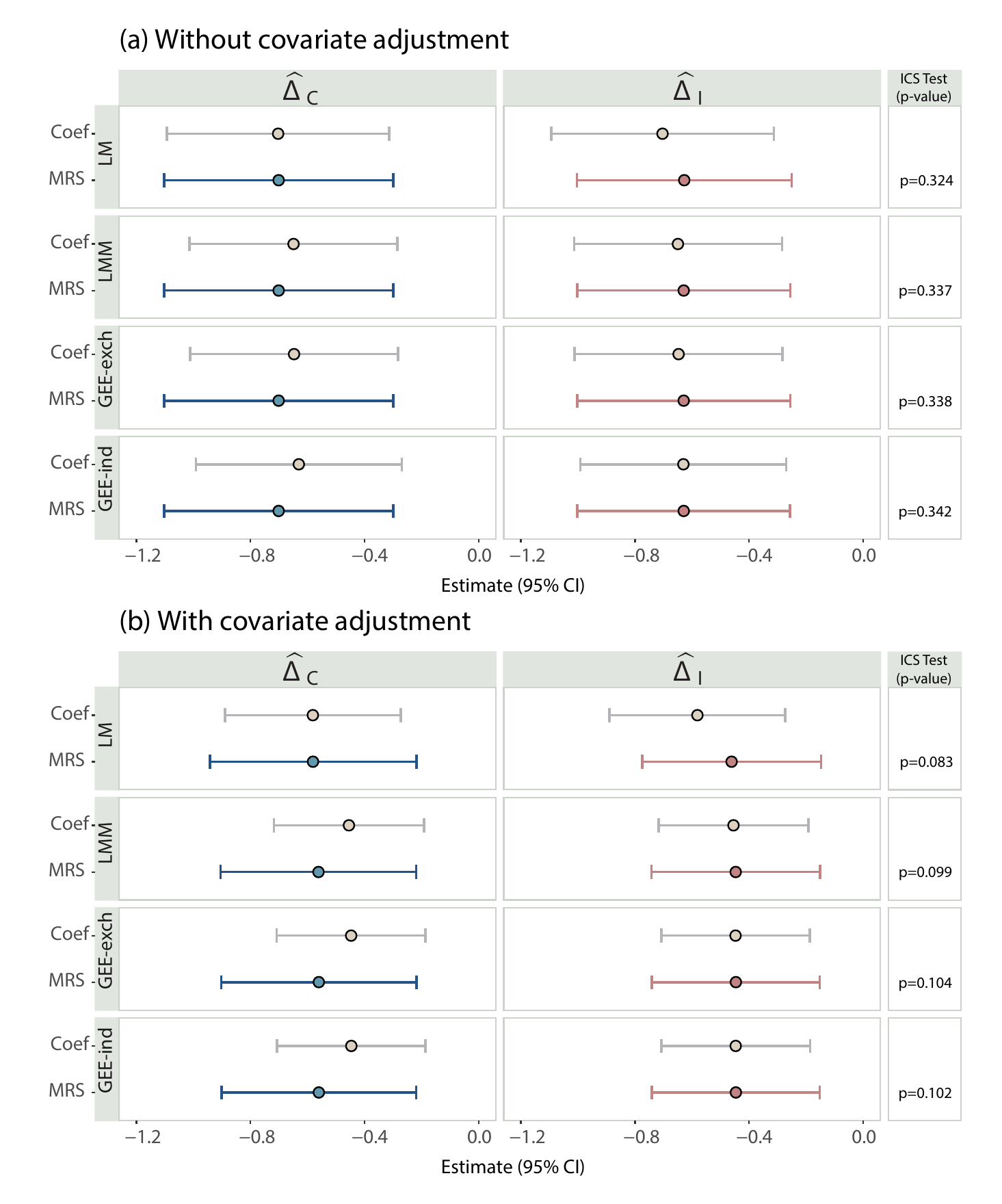}
    \caption{Analysis of the PPACT study, with the cluster-ATE and individual-ATE defined on the mean difference scale. ICS test: Informative cluster size test based on statistic \(\widehat{\Delta}_C - \widehat{\Delta}_I\). Panel (a) shows the estimates and their \(95\%\) confidence intervals without covariates adjustment, Panel (b) shows the estimates and their \(95\%\) confidence intervals with covariates adjustment.}
    \label{fig:ppact}
\end{figure}

Primary care providers (PCPs) have increasingly turned to medication as the primary treatment for chronic pain, largely due to the scarcity of alternative options. However, the rising concerns about opioid safety and efficacy emphasize the need for more effective treatment options. A promising alternative involves interdisciplinary teams within primary care settings, including behavioral specialists, nurse care managers, physical therapists, and pharmacists. The Pain Program of Active Coping and Training (PPACT) is a mixed-methods, pragmatic CRT conducted across several Kaiser Permanente regions and examines the effectiveness of an interdisciplinary approach compared to usual care for chronic pain patients on long-term opioid treatment. This trial included 106 primary care provider (PCP) and a total of 850 participants. Of these, 53 PCPs are randomized to the PPACT intervention, and the other 53 PCPs are randomized to usual care. We define the target estimand as the average potential outcome difference in PEGS score at one year between the PPACT intervention and usual care. The PEGS score (Pain, Enjoyment, General Activity) is a composite of pain severity and interference as measured by the 4-item version of the Brief Pain Inventory, ranging from 0 to 40. It evaluates pain's impact every three months over a year, considering both pain intensity and its effect on daily activities such as employment, physical activity, and sleep. We considered four different working outcome regression models in the model-robust standardization estimator: the linear model based on cluster-level means, linear mixed model based on individual-level observations, marginal model with an exchangeable working correlation, and marginal model under working independence (W1-W4, as abbreviated in the simulation study). The model coefficient effects of the intervention from the outcome regression are also reported as a comparison. Both unadjusted and covariate-adjusted analyses are considered, and the covariates adjusted for in each working model include gender, age, diagnoses of two or more chronic medical conditions (including diabetes, cardiovascular disorders, hypertension, and chronic pulmonary disease), anxiety and/or depression diagnoses, number of different pain types, average morphine milligram equivalents (MME) dose, benzodiazepine dispensed, baseline PEGS score, and satisfaction with primary care service. The variance estimation follows the same procedure used in our simulation study, without any small-sample bias correction due to the relatively large number of clusters.

The data analysis results for the PPACT trial are summarized in Figure \ref{fig:ppact}. Both cluster-level ($\Delta_{C}$) and individual-level ($\Delta_{I}$) estimators are calculated using the model-robust standardization estimator, with and without covariate adjustment. The model-based coefficient estimates are reproduced under both $\widehat{\Delta}_C$ and $\widehat{\Delta}_I$. The results show a clear pattern. Regardless of covariate adjustment, the treatment coefficient from different regression models can be somewhat different, whereas the model-robust standardization estimators tend to be much less sensitive to the choice of working outcome model. Under informative cluster size, previous studies\cite{su2021model,wang2022two,kahan2023demystifying,kahan2023estimands} and our simulations in Section \ref{s:sim;ss:continuous} have shown that, regardless of covariate adjustment, treatment effect coefficients from (W1) and (W4) are consistent to $\widehat{\Delta}_C$ and $\widehat{\Delta}_I$, respectively. This is also indirectly supported by our data analysis, as the treatment coefficient point estimates in those cases are identical to those under model-robust standardization. Furthermore, under model-robust standardization, we observe that covariate adjustment leads to some differences in the point estimates, likely due to addressing residual imbalance in finite samples. For each working regression model, covariate adjustment consistently results in narrower confidence intervals and leads to efficiency gain compared to unadjusted analyses. To conclude, based on model-robust standardization, the intervention leads to notable reduction of the PEGS score at one year, based on either the cluster-ATE ($\Delta_{C}$) or the individual-ATE ($\Delta_{I}$) estimates.

Finally, we report the p-value under a two-sided test for the informative cluster size: $\mathcal{H}_0:\Delta_{C}=\Delta_{I}$. Based on the p-values, there is not adequate evidence to reject the null hypothesis at the 0.05 level of type I error. However, the study may not be powered to detect the informative cluster sizes since we observe a small to moderate difference between estimates of $\widehat\Delta_{C}$ and $\widehat\Delta_{I}$. In addition, the p-value for testing informative cluster size is much smaller after covariate adjustment, likely because exploiting baseline information can improve the power of this test. Overall, regardless of the informativeness of the cluster size, the model-robust standardization method could be recommended as it achieves more transparent estimand-aligned analysis without overly relying on the chosen outcome regression model.

\section{Discussion}

Although there have been three decades of methods development for analyzing CRTs, a stronger call for clarifying treatment effect estimands has only emerged in the recent literature.\cite{kahan2023estimands,kahan2023demystifying} In that context, several prior studies have examined whether the model-based treatment effect coefficient can target marginal estimands under the potential outcomes framework. For example, Wang et al.\cite{wang2021mixed} proved that without informative cluster size, the treatment coefficient estimator from a linear mixed model is consistent to the average treatment effect estimand (also see Table \ref{tab_NICS} for an empirical verification) under arbitrary working model misspecification. However, that estimator generally converges to a less interpretable target parameter under informative cluster sizes.\cite{wang2022two,kahan2023demystifying} On the other hand, as two examples of model-assisted procedures, the treatment coefficient estimator from cluster-level linear regression targets the cluster-ATE, and that from the (unweighted) independence GEE targets individual-ATE, under arbitrary model misspecification.\cite{su2021model,kahan2023estimands} Although not examined by our simulations, cluster-level linear regression can also target individual-ATE and independence GEE can target individual-ATE once appropriate weights are specified.\cite{su2021model,wang2022two,kahan2023estimands,kahan2023demystifying} While these insightful results provide timely and important clarifications, a caveat is that they have only been developed for specific regression models. In addition, although the cluster-level linear regression and independence GEE are robust to arbitrary misspecification, they are much less commonly used in practice compared to mixed model regression for analyzing CRTs.\cite{fiero2016statistical}Given this discrepancy, an important objective of our paper is to demonstrate that there exists a unifying procedure---model-robust standardization---that can incorporate any regression models (including generalized linear mixed models and exchangeable GEE) previously developed for CRTs and still consistently target cluster-ATE or individual-ATE as two typical marginal estimands. This has important practical implications because one can automatically achieve estimand-aligned inference in CRTs using their preferred model specification as a working model in our procedure. We acknowledge that this work primarily focuses on aligning estimation procedures with target estimands, rather than on maximizing statistical efficiency. While covariate adjustment can generally improve efficiency relative to unadjusted estimators, such gains are not specific to the proposed model-robust standardization approach. For broader discussions on leveraging flexible outcome models---including machine learning methods---for maximizing efficiency gains in CRTs, we refer readers to recent work by Wang et al.,\cite{wang2024handling,wang2024model,wang2024asymptotic} Balzer et al.,\cite{balzer2023two} Benitez et al.,\cite{benitez2023defining} and Nugent et al.\cite{nugent2024blurring} Finally, to facilitate the application of our model-robust standardization approach in practice, we also provide an R package \texttt{MRStdCRT} that is available at \url{https://github.com/deckardt98/MRStdCRT}. A short tutorial is available in Appendix B to explain the syntax and output with the example dataset analyzed in Section \ref{sec:data}.

While the cluster jackknife variance estimator offers a computationally convenient recipe for inference under model-robust standardization, our simulation findings highlight important opportunities for future research to improve this estimator. In Section \ref{sec:simulation}, even though the jackknife variance estimator is generally close to the Monte Carlo variance and leads to nominal coverage with $m=100$ clusters, it becomes slightly conservative as the number of clusters decreases to $m=30$, and may be increasingly conservative with even fewer clusters (Web Appendix E). This pattern aligns with earlier findings in the context of simple linear regression with clustered data, as documented by Hansen.\cite{hansen2022jackknife} To some extent, our empirical results extend their observations to a broader class of outcome regression models. Although small-sample corrections for model-based inference in CRTs have been more extensively studied,\cite{mancl2001covariance,leyrat2018cluster} analogous adjustments for model-robust standardization are yet to be more systematically investigated. There remains substantial uncertainty about how best to improve confidence interval coverage for estimating cluster-ATE and individual-ATE in a small CRT. Therefore, developing principled small-sample corrections that improve the cluster jackknife represents an important and largely open area for future research. Synergistically, small-sample adjustments may also help mitigate the conservativeness observed for tests for informative cluster size.

This work stimulates the following areas that we plan to pursue in future research. First, Bayesian methods have been developed for the analysis of CRTs, but model-robust inference with Bayesian hierarchical models has not been fully elucidated. It will be interesting to develop the model-robust analogue when ${E}(\overline{Y}_{i}|A_i=a,\bm{X}_i,\bm{H}_i,N_i)$ is informed by posterior distributions of model parameters. Second, we have not study survival outcomes, but plan to further generalize the model-robust standardization estimator to accommodate covariate-dependent right censoring. Third, an increasing number of CRTs are designed with outcome data collected over several different time periods, and additionally with treatment rolled out in a staggered fashion. Specifically in stepped wedge CRTs, treatment effect estimands under the potential outcomes framework have been formalized recently, when the cluster-period size is informative\cite{chen2025model} and when the treatment effect can depend on treatment duration.\cite{wang2024achieve} It would be useful to generalize the model-robust standardization estimator such that estimand-aligned analysis of SW-CRTs can be achieved with an arbitrary outcome regression model. Fourth, while our focus is on marginal estimands defined under the potential outcomes framework, we acknowledge that these may not be the only estimands of interest. In particular, cluster-specific and individual-specific estimands, though not addressed in this work, can also be important for informing decisions at the level of specific cluster stakeholders or individual patients and are useful to study in future work. Finally, we have not explored the implications of missing outcomes within the model-robust standardization framework. A practical approach is to combine model-robust standardization with multilevel multiple imputation, consistent with existing recommendations for handling missing outcomes in model-based analyses of CRTs.\cite{turner2020properties} However, a comprehensive investigation of this strategy is left for future research.

\section{Conclusion}

When interest lies in estimating the marginal estimands such as cluster-ATE or individual-ATE, the treatment effect coefficient derived from conventional regression models may be biased under model misspecification, except in certain special scenarios. We have introduced a model-robust standardization approach for CRTs that ensures valid estimation of these marginal estimands regardless of the outcome model specification, and evaluated its performance through extensive simulations under a variety of settings. We recommend routinely employing the model-robust standardization framework in CRTs, particularly when targeting marginal ratio estimands or when the cluster size is informative, unless researchers have compelling justification to rely solely on treatment effect coefficients from specific regression models.


\section*{Acknowledgements}
Research in this article was supported by a Patient-Centered Outcomes Research Institute Award\textsuperscript{\textregistered} (PCORI\textsuperscript{\textregistered} Award ME-2022C2-27676) and the National Institute Of Allergy And Infectious Diseases of the National Institutes of Health under Award Number R00AI173395. The statements presented in this article are solely the responsibility of the authors and do not necessarily represent the official views of the National Institutes of Health or PCORI\textsuperscript{\textregistered}, its Board of Governors, or the Methodology Committee. The authors gratefully acknowledge the attendees of the 2024 Allan Donner Lecture at Western University for their insightful discussions and valuable feedback.

\section*{Supplementary Material}

The supplementary material includes technical derivations, example R code, additional simulation studies, and web tables referenced throughout the article.

    \bibliography{CRT}

\begin{thebibliography}{10}
\providecommand \doibase [0]{http://dx.doi.org/}%

\bibitem{murray1998design}
Murray DM. {\it Design and analysis of group-randomized trials}. 29.
\newblock Oxford University Press .
\newblock 1998.

\bibitem{turner2017review}
Turner EL, Li F, Gallis JA, Prague M, Murray DM. Review of recent methodological developments in group-randomized trials: part 1—design. {\it American journal of public health} 2017\string; 107(6)\string: 907--915.

\bibitem{weinfurt2017pragmatic}
Weinfurt KP, Hernandez AF, Coronado GD, et al. Pragmatic clinical trials embedded in healthcare systems: generalizable lessons from the NIH Collaboratory. {\it BMC medical research methodology} 2017\string; 17\string: 1--10.

\bibitem{kahan2023estimands}
Kahan BC, Li F, Copas AJ, Harhay MO. Estimands in cluster-randomized trials: choosing analyses that answer the right question. {\it International Journal of Epidemiology} 2023\string; 52(1)\string: 107--118.

\bibitem{kahan2023demystifying}
Kahan BC, Blette B, Harhay MO, et al. Demystifying estimands in cluster-randomised trials. {\it Statistical Methods in Medical Research} 2024\string; 33(7)\string: 1211-1232.

\bibitem{kahan2023informative}
Kahan BC, Li F, Blette B, Jairath V, Copas A, Harhay M. Informative cluster size in cluster-randomised trials: A case study from the TRIGGER trial. {\it Clinical Trials} 2023\string; 20(6)\string: 661--669.

\bibitem{donner2000design}
Donner A, Klar N, Klar NS. {\it Design and analysis of cluster randomization trials in health research}. 27.
\newblock Arnold London .
\newblock 2000.

\bibitem{campbell2012consort}
Campbell MK, Piaggio G, Elbourne DR, Altman DG. Consort 2010 statement: extension to cluster randomised trials. {\it Bmj} 2012\string; 345.

\bibitem{imai2009essential}
Imai K, King G, Nall C. The essential role of pair matching in cluster-randomized experiments, with application to the {M}exican universal health insurance evaluation. {\it Statistical Science} 2009\string; 24(1)\string: 29--53.

\bibitem{benitez2023defining}
Benitez A, Petersen ML, Laan v.~dMJ, et al. Defining and estimating effects in cluster randomized trials: A methods comparison. {\it Statistics in medicine} 2023\string; 42(19)\string: 3443--3466.

\bibitem{wang2022two}
Wang X, Turner EL, Li F, et al. Two weights make a wrong: cluster randomized trials with variable cluster sizes and heterogeneous treatment effects. {\it Contemporary clinical trials} 2022\string; 114\string: 106702.

\bibitem{su2021model}
Su F, Ding P. Model-assisted analyses of cluster-randomized experiments. {\it Journal of the Royal Statistical Society Series B: Statistical Methodology} 2021\string; 83(5)\string: 994--1015.

\bibitem{balzer2019new}
Balzer LB, Zheng W, Laan v.~dMJ, Petersen ML. A new approach to hierarchical data analysis: targeted maximum likelihood estimation for the causal effect of a cluster-level exposure. {\it Statistical Methods in Medical Research} 2019\string; 28(6)\string: 1761--1780.

\bibitem{balzer2023two}
Balzer LB, Laan v.~dM, Ayieko J, et al. Two-Stage {TMLE} to reduce bias and improve efficiency in cluster randomized trials. {\it Biostatistics} 2023\string; 24(2)\string: 502--517.

\bibitem{nugent2024blurring}
Nugent JR, Marquez C, Charlebois ED, Abbott R, Balzer LB. Blurring cluster randomized trials and observational studies: Two-Stage TMLE for subsampling, missingness, and few independent units. {\it Biostatistics} 2024\string; 25(3)\string: 599--616.

\bibitem{wang2024handling}
Wang B, Li F, Wang R. Handling incomplete outcomes and covariates in cluster-randomized trials: doubly-robust estimation, efficiency considerations, and sensitivity analysis. {\it arXiv preprint arXiv:2401.11278} 2024.

\bibitem{wang2024model}
Wang B, Park C, Small DS, Li F. Model-robust and efficient covariate adjustment for cluster-randomized experiments. {\it Journal of the American Statistical Association} 2024\string: 1--13.

\bibitem{fiero2016statistical}
Fiero MH, Huang S, Oren E, Bell ML. Statistical analysis and handling of missing data in cluster randomized trials: a systematic review. {\it Trials} 2016\string; 17(1)\string: 1--10.

\bibitem{offorha2022statistical}
Offorha BC, Walters SJ, Jacques RM. Statistical analysis of publicly funded cluster randomised controlled trials: a review of the National Institute for Health Research Journals Library. {\it Trials} 2022\string; 23(1)\string: 115.

\bibitem{hemming2023commentary}
Hemming K, Taljaard M. Commentary: Estimands in cluster trials: thinking carefully about the target of inferenceand the consequences for analysis choice. {\it International Journal of Epidemiology} 2023\string; 52(1)\string: 116--118.

\bibitem{wang2021mixed}
Wang B, Harhay MO, Small DS, Morris TP, Li F. On the mixed-model analysis of covariance in cluster-randomized trials. {\it arXiv preprint arXiv:2112.00832} 2021.

\bibitem{lewsey2004comparing}
Lewsey J. Comparing completely and stratified randomized designs in cluster randomized trials when the stratifying factor is cluster size: a simulation study. {\it Statistics in medicine} 2004\string; 23(6)\string: 897--905.

\bibitem{li2016evaluation}
Li F, Lokhnygina Y, Murray DM, Heagerty PJ, DeLong ER. An evaluation of constrained randomization for the design and analysis of group-randomized trials. {\it Statistics in medicine} 2016\string; 35(10)\string: 1565--1579.

\bibitem{li2017evaluation}
Li F, Turner EL, Heagerty PJ, Murray DM, Vollmer WM, DeLong ER. An evaluation of constrained randomization for the design and analysis of group-randomized trials with binary outcomes. {\it Statistics in medicine} 2017\string; 36(24)\string: 3791--3806.

\bibitem{zou2009assessment}
Zou G. Assessment of risks by predicting counterfactuals. {\it Statistics in medicine} 2009\string; 28(30)\string: 3761--3781.

\bibitem{van2000asymptotic}
{Van der Vaart} AW. {\it Asymptotic Statistics}.
\newblock Cambridge university press .
\newblock 2000.

\bibitem{hines2022demystifying}
Hines O, Dukes O, Diaz-Ordaz K, Vansteelandt S. Demystifying statistical learning based on efficient influence functions. {\it The American Statistician} 2022\string; 76(3)\string: 292--304.

\bibitem{zeng2021propensity}
Zeng S, Li F, Wang R, Li F. Propensity score weighting for covariate adjustment in randomized clinical trials. {\it Statistics in medicine} 2021\string; 40(4)\string: 842--858.

\bibitem{wang2023model}
Wang B, Susukida R, Mojtabai R, Amin-Esmaeili M, Rosenblum M. Model-robust inference for clinical trials that improve precision by stratified randomization and covariate adjustment. {\it Journal of the American Statistical Association} 2023\string; 118(542)\string: 1152--1163.

\bibitem{johnson2015recommendations}
Johnson JL, Kreidler SM, Catellier DJ, Murray DM, Muller KE, Glueck DH. Recommendations for choosing an analysis method that controls Type I error for unbalanced cluster sample designs with Gaussian outcomes. {\it Statistics in Medicine} 2015\string; 34(27)\string: 3531--3545.

\bibitem{hayes2017cluster}
Hayes RJ, Moulton LH. {\it Cluster randomised trials}.
\newblock Chapman and Hall/CRC .
\newblock 2017.

\bibitem{raudenbush1997statistical}
Raudenbush SW. Statistical analysis and optimal design for cluster randomized trials.. {\it Psychological Methods} 1997\string; 2(2)\string: 173.

\bibitem{tong2022accounting}
Tong G, Esserman D, Li F. Accounting for unequal cluster sizes in designing cluster randomized trials to detect treatment effect heterogeneity. {\it Statistics in Medicine} 2022\string; 41(8)\string: 1376--1396.

\bibitem{crouch1990evaluation}
Crouch EA, Spiegelman D. The Evaluation of Integrals of the form $\int_{\infty}^{+\infty} f(t)\exp(-t^2)dt$: Application to Logistic-Normal Models. {\it Journal of the American Statistical Association} 1990\string; 85(410)\string: 464--469.

\bibitem{hedeker2018note}
Hedeker D, Toit dSH, Demirtas H, Gibbons RD. A note on marginalization of regression parameters from mixed models of binary outcomes. {\it Biometrics} 2018\string; 74(1)\string: 354--361.

\bibitem{ritz2004equivalence}
Ritz J, Spiegelman D. Equivalence of conditional and marginal regression models for clustered and longitudinal data. {\it Statistical Methods in Medical Research} 2004\string; 13(4)\string: 309--323.

\bibitem{liang1986longitudinal}
Liang KY, Zeger SL. Longitudinal data analysis using generalized linear models. {\it Biometrika} 1986\string; 73(1)\string: 13--22.

\bibitem{preisser2003integrated}
Preisser JS, Young ML, Zaccaro DJ, Wolfson M. An integrated population-averaged approach to the design, analysis and sample size determination of cluster-unit trials. {\it Statistics in Medicine} 2003\string; 22(8)\string: 1235--1254.

\bibitem{li2021sample}
Li F, Tong G. Sample size and power considerations for cluster randomized trials with count outcomes subject to right truncation. {\it Biometrical Journal} 2021\string; 63(5)\string: 1052--1071.

\bibitem{li2021sample_bin}
Li F, Tong G. Sample size estimation for modified Poisson analysis of cluster randomized trials with a binary outcome. {\it Statistical Methods in Medical Research} 2021\string; 30(5)\string: 1288--1305.

\bibitem{preisser2008finite}
Preisser JS, Lu B, Qaqish BF. Finite sample adjustments in estimating equations and covariance estimators for intracluster correlations. {\it Statistics in medicine} 2008\string; 27(27)\string: 5764--5785.

\bibitem{westgate2012bias}
Westgate PM. A bias-corrected covariance estimate for improved inference with quadratic inference functions. {\it Statistics in Medicine} 2012\string; 31(29)\string: 4003--4022.

\bibitem{yu2020evaluation}
Yu H, Li F, Turner EL. An evaluation of quadratic inference functions for estimating intervention effects in cluster randomized trials. {\it Contemporary Clinical Trials Communications} 2020\string; 19\string: 100605.

\bibitem{heagerty2000marginalized}
Heagerty PJ, Zeger SL. Marginalized multilevel models and likelihood inference (with comments and a rejoinder by the authors). {\it Statistical Science} 2000\string; 15(1)\string: 1--26.

\bibitem{hansen2022jackknife}
Hansen BE. Jackknife standard errors for clustered regression. {\it University of Wisconsin} 2022.

\bibitem{mancl2001covariance}
Mancl LA, DeRouen TA. A covariance estimator for GEE with improved small-sample properties. {\it Biometrics} 2001\string; 57(1)\string: 126--134.

\bibitem{ouyang2024maintaining}
Ouyang Y, Taljaard M, Forbes AB, Li F. Maintaining the validity of inference from linear mixed models in stepped-wedge cluster randomized trials under misspecified random-effects structures. {\it Statistical Methods in Medical Research} 2024\string: 09622802241248382.

\bibitem{morris2019using}
Morris TP, White IR, Crowther MJ. Using simulation studies to evaluate statistical methods. {\it Statistics in Medicine} 2019\string; 38(11)\string: 2074--2102.

\bibitem{leyrat2018cluster}
Leyrat C, Morgan KE, Leurent B, Kahan BC. Cluster randomized trials with a small number of clusters: which analyses should be used?. {\it International Journal of Epidemiology} 2018\string; 47(1)\string: 321--331.

\bibitem{wang2024asymptotic}
Wang B, Li F. Asymptotic inference with flexible covariate adjustment under rerandomization and stratified rerandomization. {\it arXiv preprint arXiv:2406.02834} 2024.

\bibitem{chen2025model}
Chen X, Li F. Model-assisted analysis of covariance estimators for stepped wedge cluster randomized experiments. {\it Scandinavian Journal of Statistics} 2025\string; 52(1)\string: 416--446.

\bibitem{wang2024achieve}
Wang B, Wang X, Li F. How to achieve model-robust inference in stepped wedge trials with model-based methods?. {\it Biometrics} 2024\string; 80(4)\string: ujae123.

\bibitem{turner2020properties}
Turner EL, Yao L, Li F, Prague M. Properties and pitfalls of weighting as an alternative to multilevel multiple imputation in cluster randomized trials with missing binary outcomes under covariate-dependent missingness. {\it Statistical Methods in Medical Research} 2020\string; 29(5)\string: 1338--1353.

\end{thebibliography}

\end{document}